\RequirePackage{ifpdf}
\documentclass[11pt,letterpaper]{JHEP3}
\usepackage{epsfig}
\usepackage[latin1]{inputenc}
\usepackage{bbold,amsfonts}
\usepackage{graphicx}
\usepackage{amssymb,amsmath}
\usepackage{fancybox,framed}
\usepackage{dsfont}
\usepackage{mathtools}
\usepackage{braket}
\usepackage{slashed}
\usepackage{rotating}

\author{Lorenzo~Bianchi$^{a,b}$, Marco S. Bianchi$^{c}$\\\\
$^{a}$ Institut f\"ur Physik,
Humboldt-Universit\"at
zu Berlin\\ Zum Gro\ss en Windkanal 6, 12489 Berlin, Germany
\\
$^{b}$ II. Institut f\"ur Theoretische Physik,
Universit\"at Hamburg\\
Luruper Chaussee 149,
22761 Hamburg, Germany
\\
$^{c}$
Centre for Research in String Theory,
School of Physics and Astronomy\\
Queen Mary University of London,
Mile End Road, London E1 4NS, UK \\
\qquad\\
E-mail: \email{lorenzo.bianchi@desy.de, m.s.bianchi@qmul.ac.uk}
}

\abstract{We compute the one-loop S-matrix for the light bosonic excitations of the GKP string at strong coupling. These correspond, on the gauge theory side, to gluon insertions in the GKP vacuum. We perform the calculation by Feynman diagrams in the worldsheet theory and
we compare the result to the integrability prediction, finding perfect agreement for the scheme independent part. For scheme dependent rational terms we test different schemes and find that a recent proposal reproduces exactly the integrability prediction.
}

\preprint{November 2015\\DESY 15-200 \\QMUL-PH-15-19
}

\title{On the scattering of gluons in the GKP string
}


\csname @addtoreset\endcsname{equation}{section}

\def\bseq{\begin{subequation}}  
\def\eseq{\end{subequation}}
\def\bsea{\begin{subeqnarray}}  
\def\esea{\end{subeqnarray}}

\newcommand{\beq}{\begin{equation}}
\newcommand{\bea}{\begin{eqnarray}}
\newcommand{\eea}{\end{eqnarray}}
\newcommand{\eeq}{\end{equation}}

\renewcommand{\a}{\alpha}
\renewcommand{\b}{\beta}

\renewcommand{\d}{\delta}

\newcommand{\g}{\gamma}

\newcommand{\D}{\Delta}

\renewcommand{\P}{\Pi}

\newcommand{\ppp}{\text{p}}

\def\beq{\begin{equation}}
\def\eeq{\end{equation}}
\def\bea{\begin{eqnarray}}
\def\eea{\end{eqnarray}}

\def\a{\alpha}
\def\b{\beta}
\def\g{\gamma}

\def\d{\delta}

\def\th{\theta}

\def\D{\Delta}

\declareslashed{}{\backslash}{0}{0}{\Omega}
\declareslashed{}{\backslash}{0}{0}{O}
\begin{document}

\allowdisplaybreaks

\section{Introduction}

A paramount advance in the study of scattering amplitudes of planar ${\cal N}=4$ SYM has been fostered by insights from the AdS/CFT correspondence \cite{Maldacena:1997re} and integrability \cite{Minahan:2002ve,Beisert:2003yb}.
In planar ${\cal N}=4$ SYM (super)amplitudes are dual to null polygonal (super)Wilson loops \cite{Alday:2007hr,Drummond:2007au,Drummond:2007cf,Brandhuber:2007yx,CaronHuot:2010ek,Mason:2010yk}.
The expectation value of such operators can be formally reconstructed in terms of a sum over excitations of the color flux-tube supported by the loop \cite{Alday:2010ku,Basso:2013vsa}.

The philosophy parallels that of an OPE expansion for the $n$-point function of local operators in a conformal field theory, but is instead applied to a non-local Wilson loop operator whose contour is a cusped polygon with $n$ edges. The latter is geometrically decomposed into a sequence of $(n-5)$ elementary pentagons, each parameterised by three variables associated to the energy, momentum and angular momentum of the excitations, and dual to the $3(n-5)$ conformal cross-ratios describing the scattering amplitude.
The excitations propagate through pentagons with their dispersion relation \cite{Basso:2010in}. The transition between two adjacent pentagons is governed by the intricate dynamics of the flux-tube theory and is captured by an object dubbed the pentagon transition.
Remarkably, a set of axioms determines such amplitudes and in particular relates them to the scattering elements of the flux-tube theory.
The latter can be identified as an operator of large spin \cite{Alday:2007mf} amenable of an integrable spin chain interpretation, or, at strong coupling, as the dual excited GKP string.
Integrability of ${\cal N}=4$ SYM then allows to derive exact results for the spectrum of the excitations and their scattering factors, valid at any coupling.
\bigskip

The OPE program has been intensively studied at weak coupling \cite{Basso:2013aha,Belitsky:2014rba,Basso:2014koa,Belitsky:2014sla,Basso:2014nra,Belitsky:2014lta,Belitsky:2015efa,Basso:2014hfa,Basso:2015rta}, where several results are available to rather high order in perturbation theory \cite{Dixon:2011nj,Dixon:2014iba,Golden:2014xqa,Golden:2014xqf,Golden:2014pua,Drummond:2014ffa}.
The computation of amplitudes at strong coupling is more elusive, but certainly challenging and fascinating (and even more interesting is that at finite coupling \cite{Basso:2015uxa}).
Contrary to the weak coupling expansion, where the number of contributions of the various excitations and bound states thereof can be truncated at a given perturbative order, the strong coupling expansion entails summing over an infinite series of excitations.

Recent publications \cite{Basso:2013vsa,Basso:2014koa,Basso:2014jfa,Fioravanti:2015dma,Belitsky:2015qla} have been paving the road to two major progresses in this quest. The first would be the extension of the leading order results for $n$-point MHV amplitudes, whose minimal area problem is solved via the Termodynamic Bethe Ansatz \cite{Alday:2007hr,Alday:2009dv,Alday:2010vh}, to NMHV amplitudes \cite{Belitsky:2015qla}.
The second would be the determination of MHV amplitudes at next-to-leading order at strong coupling.
This task requires the knowledge of the higher order corrections to the pentagon transitions at strong coupling.
This in turn can be achieved using their conjectured relation in terms of S-matrix elements of the flux-tube excitations.
Integrability of the flux-tube theory, the GKP string model \cite{Gubser:2002tv,Frolov:2002av}, allows to determine the S-matrix from a set of Bethe equations. Nevertheless, performing their expansion at strong coupling is non-trivial. Results for scattering of fermion, gauge excitations and bound states thereof in the strong perturbative regime have been recently derived in \cite{Fioravanti:2013eia,Fioravanti:2015dma,Belitsky:2015qla}.
\bigskip

The S-matrix of GKP excitations at strong coupling can be studied more traditionally by means of perturbation theory within the worldsheet theory describing the GKP string.
In particular the light-cone gauge fixed Metsaev-Tseytlin Lagrangian for the $AdS_5\times S^5$ superstring \cite{Metsaev:1998it,Metsaev:2000yf,Metsaev:2000yu}, expanded in fluctuations around the GKP (or equivalently the null cusp) vacuum \cite{Giombi:2009gd}, has been proven a powerful starting point for perturbative computation.
It allowed to compute the cusp anomalous dimension of ${\cal N}=4$ SYM at two loops at strong coupling \cite{Giombi:2009gd} and (with some caveats pointed out in \cite{Zarembo:2011ag}) the one-loop dispersion relations \cite{Giombi:2010bj} of the GKP excitations\footnote{See \cite{Bianchi:2014ada,Bianchi:2015laa} for similar results in ABJM theory.}.
Recently it has been employed to address the computation of scattering amplitudes of GKP excitations.
In particular, at leading order it reproduces the strong coupling results for several of the GKP string S-matrix elements \cite{Bianchi:2015iza}. However, some turn out to produce inconsistent results. This is due to the deeply non-perturbative dynamics of the massless scalar excitations, which the perturbative expansion is not able to capture, along the lines of \cite{Zarembo:2011ag}. This can be an issue for loop computation, where the massless scalars can trigger IR divergences which invalidate perturbation theory.
\bigskip

In this paper we focus on scattering between gauge excitations and compute their amplitude at next-to-leading order at strong coupling via perturbation theory in the worldsheet model.
The restriction to the gluonic sector is motivated by the fact that these are arguably the technically simplest amplitudes to compute via a Feynman diagram approach, as the tree-level computation already suggests.
Moreover the aforementioned subtleties associated to the non-perturbative dynamics of scalars are not a concern for this computation.
Indeed, by inspecting the possible Feynman diagrams, we ascertain that no potentially dangerous coupling to the massless scalars is involved.
This does not exclude that IR divergences at higher orders can appear spoiling the validity of our perturbative approach.
Nevertheless, the theorems of \cite{Elitzur:1978ww,David:1980gi} suggest that no such IR divergences should appear in an $SO(6)$ invariant quantity, as the scattering amplitude of gauge excitations.  

After listing all relevant Feynman diagrams for the computation, we evaluate them using the Feynman rules and find a set of tensor integrals to be evaluated.
Some of them are divergent and a regularization is needed in intermediate steps. 
The issue of UV regularization has been often discussed and analyzed in the related context of the next-to-leading order perturbative computation of the worldsheet S-matrices for near-BMN string sigma models \cite{Bianchi:2013nra,Engelund:2013fja,Roiban:2014cia}, which has recently made significant progress by both standard techniques \cite{Sundin:2013ypa,Abbott:2013kka,Sundin:2014ema,Sundin:2014sfa,Sundin:2015uva,Roiban:2014cia} and unitarity methods \cite{Bianchi:2013nra,Engelund:2013fja,Bianchi:2014rfa,Engelund:2014pla}. In particular, though the expected UV finiteness of the one-loop result has been ascertained, different regularization schemes produce different results. This is not surprising since the derivation of an exact S-matrix is based on symmetry considerations and it is crucial to find a regularization which preserves such symmetries. 

On general grounds, integrability as well as other classical symmetries broken by the regulator can
be restored by the addition to the S-matrix of matrix elements of finite local counterterms in
the effective action (see, e.g., \cite{deVega:1981ka,deVega:1982sh,Hoare:2010fb} for the example of complex Sine-Gordon theory). Nevertheless, it is an interesting and open question to find a regularization procedure that preserves all the symmetries of the
worldsheet theory. 
At one loop, Roiban, Sundin, Tseytlin and Wulff (RSTW) \cite{Roiban:2014cia} showed that the use of algebraic identities in $d=2$ provides an effective symmetry-preserving regulator for the integral reduction in the near-BMN theory.

The computation of the one-loop S-matrix for GKP excitations that we perform in this paper constitutes a non-trivial testing ground for the RSTW regularization scheme. 
Our setting is complicated by the presence of box and triangle topologies induced by three-point interactions. Moreover, at a difference with respect to the near-BMN computation of \cite{Roiban:2014cia}, the GKP gluon dispersion relation is already corrected at one loop, inducing a contribution at next-to-leading order. 
For completeness, we perform tensor reduction via three different methods and compare them.
These are Passarino-Veltman reduction in $d$ dimensions, in strictly 2 dimensions and finally the RSTW reduction procedure.
\bigskip

After performing tensor reduction, we are able to determine the next-to-leading order scattering amplitudes for gauge excitations.
We compare them to the integrability predictions and find that the scheme independent part agrees completely. This is the main result of the paper.
Furthermore, by comparing scheme dependent terms, we find that the RSTW scheme reproduces exactly the same expression as from integrability. 
\bigskip

The plan of the paper is as follows.
In Section \ref{sec:result} we summarize our results omitting all the technical details of the computation or of the expansion of the exact result.
After reviewing the form of the worldsheet Lagrangian and extracting the Feynman rules in Section \ref{sec:Lagrangian} we proceed in Section \ref{sec:tree}, recalling the worldsheet computation of scattering factors of gauge excitations at leading order, which was carried out in \cite{Bianchi:2015iza}.

In Section \ref{sec:tensor} we provide details on the most technical part of the computation, namely the reduction of tensor integrals emerging from Feynman diagrams to scalar bubble integrals. In particular we analyse and compare different approaches to tensor reduction.

We then turn to the computation of the one-loop scattering factors for same helicity gluons and opposite helicity gluons in forward and backward kinematics, in Sections \ref{sec:samehelicity}, \ref{sec:opphelicityf} and \ref{sec:opphelicityb} respectively.
For each we list and compute the Feynman diagrams at one loop. We reduce the tensor integrals and express the result in terms of scalar bubbles.
 
We finalise the one-loop computation computing two final ingredients. First, in Section \ref{sec:external}, we derive the external legs corrections contributing via the LSZ formalism.
Second, in Section \ref{sec:treedisp}, we determine the $g^{-2}$ corrections coming from evaluating the leading order result at a quantum corrected value of the particle energies in terms of the spatial momentum.

The sum of all these contributions gives the final scattering factors at next-to-leading order.
In Section \ref{sec:final} we compare these results with the prediction form integrability and find agreement. 

We provide several technical details in a series of appendices.

\section{Summary of the results}\label{sec:result}

The GKP string worldsheet theory is a classically integrable model. Assuming integrability at the quantum level, the asymptotic Bethe ansatz \cite{Beisert:2006ez} determines scattering between its excitations at any coupling $g\equiv \frac{\sqrt{\lambda}}{4\pi}$, where $\lambda$ is the ${\cal N}=4$ SYM 't Hooft coupling.
In particular the Bethe equations are amenable of a perturbative expansion at strong coupling, which allows to determine closed analytic expressions for scattering factors. 
In this paper we focus on scattering of gluon excitations up to next-to-leading order (namely order $g^{-2}$).

Here we summarize the final result from the integrability prediction, omitting all the technical details of the derivations.
These are based on the expansion of the exact result carried out in \cite{Belitsky:2015qla} and are summarized in Appendix \ref{app:integrability}.
Eventually, our perturbative computation using the light-cone gauge-fixed string sigma model precisely agrees with the integrability prediction,therefore this section also provides a synthesis of the results obtained via sigma model perturbation theory.

We express the final result in terms of the spatial components $\ppp_1$ and $\ppp_2$ of the two-momenta $p_1$ and $p_2$,  
\begin{equation}
p_i = \left( e_i, \ppp_i \right)
\end{equation}
parametrizing them by hyperbolic rapidities
\begin{equation}\label{eq:hyperbolic}
\ppp_i=\sqrt{2} \sinh \theta_i \qquad e_i = i \sqrt{2} \cosh{\theta_i} + \mathcal{O}(g^{-1})
\end{equation}
where the energy takes imaginary values for real rapidities, since we are dealing with a Euclidean worldsheet.
We stress that, since the dispersion relation is non-relativistic, the energy $e_i$ receives quantum corrections. In order to take into account the one-loop effect of those additional contributions one has to correct the energy factors and the Bethe rapidities appearing in the tree-level result, in the perturbative and integrability description, respectively. On the other hand in the one-loop terms we can safely assume a relativistic dispersion relation since the corresponding corrections would affect the results starting from two loops.

We start the summary with the same helicity amplitude
\begin{equation}\label{eq:predictionsame}
S_{gg}(\th_1,\th_2) = 1 + \frac{i}{g}\, S_{gg}^{(0)}(\th_1,\th_2)
+ \frac{i}{g^2}\, S_{gg}^{(1)}(\th_1,\th_2) + {\cal O}(g^{-3})
\end{equation}
with 
\begin{equation}\label{eq:predictionsametree}
 S_{gg}^{(0)}(\th_1,\th_2)=\frac{\cosh \left(\theta _1-\theta _2\right)+1}{2 \left(\tanh 2 \theta _1 - \tanh 2 \theta _2\right)}
\end{equation}
The one-loop expressions are lengthy and we can organize their contribution splitting it into terms, according to which transcendental number they would be proportional after collecting a factor $\frac{1}{4\pi}$.
For same helicity gluons the one-loop piece consists of
\begin{itemize}
\item an imaginary (for real rapidities) term corresponding to the square of the tree-level amplitude (as expected by unitarity arguments)
\begin{equation}\label{eq:predictionsamereal}
S_{gg}^{(1)}(\th_1,\th_2)\Big|_{i} = \frac{i}{2} \left[S_{gg}^{(0)}(\th_1,\th_2)\right]^2
\end{equation}
\item a term proportional to $\log 2$
\begin{equation}
S_{gg}^{(1)}(\th_1,\th_2)\Big|_{\log 2} = \frac{3 \log 2}{4\pi} S_{gg}^{(0)}(\th_1,\th_2)
\end{equation}
\item a term proportional to $\pi$
\begin{align}
S_{gg}^{(1)}(\th_1,\th_2)\Big|_{\pi} &= -
\frac{\cosh 2 \theta _1 \cosh 2 \theta _2}{32\, \sinh 2\left(\theta _1-\theta _2\right)} 
\left[
\cosh \left(\theta _1-\theta _2\right) \left(\cosh ^2 2 \theta _1+\cosh ^2 2 \theta _2\right) + \right.\\&
+\frac{\cosh 2 \theta _1 \cosh 2 \theta _2 \left(1-\cosh \left(\theta _1+\theta _2\right)\right)}{\cosh \theta _1\cosh \theta _2 \cosh\left(\theta _1-\theta _2\right)} +  \nonumber\\& \left.
-\frac{\sinh \left(\theta _1-\theta _2\right)}{\cosh \theta _1 \cosh \theta _2} \left(\sinh\theta _1 \cosh ^2 2 \theta _1 \cosh \theta _2 - \sinh \theta _2 \cosh \theta _1 \cosh ^2 2 \theta _2\right)\right]\nonumber
\end{align}
\end{itemize}
No other rational terms are present in the integrability prediction, in particular no algebraic numbers appear.

Analogously, for opposite helicities
\begin{equation}\label{eq:predictionopposite}
S_{gg^*}(\th_1,\th_2) = 1 + \frac{i}{g}\, S_{gg^*}^{(0)}(\th_1,\th_2)
+ \frac{i}{g^2}\, S_{gg^*}^{(1)}(\th_1,\th_2) + {\cal O}(g^{-3})
\end{equation}
with 
\begin{equation}\label{eq:predictionoppositetree}
 S_{gg^*}^{(0)}(\th_1,\th_2)=\frac{\cosh \left(\theta _1-\theta _2\right)-1}{2 \left(\tanh 2 \theta _1 - \tanh 2 \theta _2\right)}
\end{equation}
The one-loop contribution splits into
\begin{itemize}
\item a real part given by the square of the tree-level amplitude
\begin{equation}\label{eq:predictionoppreal}
S_{gg^*}^{(1)}(\th_1,\th_2)\Big|_{real} =\frac{i}{2} \left[S_{gg^*}^{(0)}(\th_1,\th_2)\right]^2
\end{equation}
\item a term proportional to $\log 2$
\begin{equation}
S_{gg^*}^{(1)}(\th_1,\th_2)\Big|_{\log 2} = \frac{3 \log 2}{4\pi} S_{gg^*}^{(0)}(\th_1,\th_2)
\end{equation}
\item a term proportional to $\pi$
\begin{align}
S_{gg^*}^{(1)}(\th_1,\th_2)\Big|_{\pi} &= 
-  \frac{\cosh 2 \theta _1 \cosh 2 \theta _2}{32\, \sinh 2\left(\theta _1-\theta _2\right)} 
\left[
\cosh \left(\theta _1-\theta _2\right) \left(\cosh ^2 2 \theta _1+\cosh ^2 2 \theta _2\right) + \right.\\&
-\frac{\cosh 2 \theta _1 \cosh 2 \theta _2 \left(1+\cosh \left(\theta _1+\theta _2\right)\right)}{\cosh \theta _1\cosh \theta _2 \cosh\left(\theta _1-\theta _2\right)} +  \nonumber\\& \left.
-\frac{\sinh \left(\theta _1-\theta _2\right)}{\cosh \theta _1 \cosh \theta _2} \left(\sinh\theta _1 \cosh ^2 2 \theta _1 \cosh \theta _2 - \sinh \theta _2 \cosh \theta _1 \cosh ^2 2 \theta _2\right)\right]\nonumber
\end{align}
\end{itemize}
Backward scattering of gluons of different helicity is absent and the S-matrix is thus reflectionless.

\section{Lagrangian and Feynman rules}\label{sec:Lagrangian}

In this section we briefly introduce the worldsheet theory for the GKP string, its spectrum of excitations, the Feynman rules and the basic ingredients for the perturbative computation of its S-matrix.
The GKP string can be described equivalently by the light-cone gauge euclidean Metsaev-Tseytlin Lagrangian for the $AdS_5 \times S^5$ sigma model, expanded in fluctuations about the null cusp vacuum \cite{Giombi:2010bj,Zarembo:2011ag}. The action reads
\begin{equation}\label{eq:action}
S = \frac{T}{2}\int dt \int^\infty_{-\infty} ds\ {\cal L}\qquad\qquad T\equiv \frac{\sqrt{\lambda}}{2\pi}
\end{equation}
where the string tension $T$ depends on the ${\cal N}=4$ 't Hooft coupling $\lambda$ and
\begin{align}\label{eq:lagrangian}
{\cal L}  &=
\big|\partial_t x + x \big|^2 +
\frac{1}{z^4} \big| \partial_s x - x \big|^2 + \Big( \partial_t z^M + z^M +
\frac{i}{z^2} \psi^{\dagger}_i \P_{+} (\rho^{MN}){}^i{}_j \psi^j  z_N \Big)^2
+ \nonumber\\
& + \frac{1}{z^{4}} \Big(\partial_s z^M - z^M \Big)^2 + 2\, i\, \psi^{\dagger}_i \partial_t \psi^i - \frac{1}{z^{2}} \Big(\psi^{\dagger}_i \P_{+} \psi^i\Big)^2 + \nonumber\\& + \frac{2i}{z^3}\, \Bigl[-\bar\psi_i \P_{+} (\rho^{\dagger}_6)^{ik} (\rho^M)_{kj} z^M \D_s \psi^j
- \frac{i}{z} (\psi^i)^T \P_{+} (\rho^M)_{ij} z^M \psi^j \D_s x + \nonumber\\& ~~~~~~~~
+ \psi^{\dagger}_i \P_{+} (\rho^\dagger_M)^{ik} z^M (\rho^6)_{kj} \D_s \psi^j
+ \frac{i}{z} \psi^{\dagger}_i \P_{+} (\rho^{\dagger}_M)^{ij} z^M (\psi^{\dagger})_j \D_s x^*\Bigr]
\end{align}
with
\begin{eqnarray}
& z = e^{\phi}\,, \qquad\qquad
z^M = e^{\phi} u^M\,, \qquad\qquad M=1,\dots 6 & \nonumber\\
& \displaystyle u^{a} = \frac{y^{a}}{1+\frac{1}{4}y^2}\,, \qquad\qquad
u^{6} = \frac{1-\frac{1}{4}y^2}{1+\frac{1}{4}y^2}\,, \qquad\qquad y^2\equiv \sum_{a=1}^5 (y^a)^2\,, \quad\qquad a=1,...,5 &
\end{eqnarray}
and $\D_s \equiv \partial_s-1$. The $\rho^{M}_{ij} $ matrices are the off-diagonal blocks of 6d gamma matrices in chiral representation and $(\rho^{MN})_i^{\phantom{i}j} = (\rho^{[M}\rho^{\dagger N]})_i^{\phantom{i}j}$ and $(\rho^{MN})^i_{\phantom{i}j} = ( \rho^{\dagger [M}\rho^{N]})^i_{\phantom{i}j}$ are the $SO(6)$ Lorentz matrices.

The gamma matrices are defined as
\begin{equation}
\g^t = -\sigma_1 \qquad\qquad \g^s = \sigma_3
\end{equation}
and $\bar\psi \equiv \psi^{\dagger}\g^t$. The projectors appearing in the Lagrangian are defined $\P_{\pm} \equiv \frac12 \left( \mathbb{1} \pm \g^s \right)$, where $\mathbb{1}$ is the $2\times 2$ identity matrix.

The spectrum of the fluctuations is derived by expanding in the fields to second order 
\begin{equation}
{\cal L}_2  = \partial_{\alpha} \phi\, \partial_{\alpha} \phi +4\,\phi^2 +
\partial_{\alpha} x\, \partial_{\alpha} x^*
+2\, x\, x^{*}
+\partial_{\alpha}y^a\partial_{\alpha}y^a
+ 2\,i\, \bar \psi_i \left(\slashed{\partial} + \mathbb{1} \right)\psi^i
\label{eq:quadratic}
\end{equation}
The bosonic sector consists of a mass $\sqrt{2}$ complex scalar $x$, a mass 2 scalar $\phi$ and 5 massless scalars $y^a$, $a=1,\dots 5$.
The fermionic sector consists of 8 mass 1 Dirac fermions $\psi^i$, $i=1,\dots 4$, transforming in the $\bf{4}$ representation of $SU(4)$.
In this paper we focus on scattering of the $x$ and $x^*$ particles, which are interpreted as the insertion of a positive and negative helicity gluon in the GKP vacuum.

The Feynman rules for the Lagrangian \eqref{eq:lagrangian} follow.
The propagators extracted from \eqref{eq:quadratic} read
\begin{align}\label{eq:props}
\langle x(p)x^*(-p) \rangle &= \raisebox{-1mm}{\includegraphics[width=3cm]{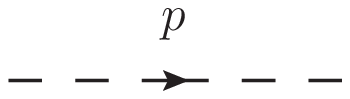}} = \frac{1}{2g}\, \frac{2}{p^2+2}\nonumber\\
\langle \phi(p)\phi(-p) \rangle &= \raisebox{-1mm}{\includegraphics[width=3cm]{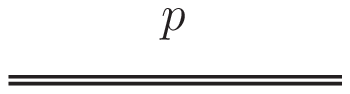}} = \frac{1}{2g}\, \frac{1}{p^2+4}\nonumber\\
\langle y^a(p)y^b(-p) \rangle &= \raisebox{-1mm}{\includegraphics[width=3cm]{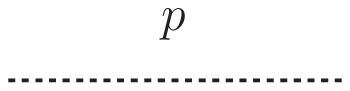}} = \frac{1}{2g}\, \frac{\delta^{ab}}{p^2}\nonumber\\
\langle \psi^i(p)\bar\psi_j(-p) \rangle &= \raisebox{-1mm}{\includegraphics[width=3cm]{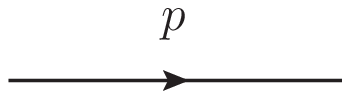}} = \frac{1}{2g}\, i\, \frac{i\slashed{p}-\mathbb{1}}{p^2+1}\, \delta^i_{\phantom{i}j}
\end{align}
Interaction vertices are straightforwardly inferred from the action. 
For the one-loop computation we carry out in this paper we have to expand some vertices involving $x$ fields up to order five.
We list them for completeness in the Appendix \ref{app:lagr_exp}.
Taking into account the non-standard normalization of the action, the prescription for the correct coefficient of vertices amounts to multiplying each by a factor $-\frac12$. 

We now turn to kinematics.
Scattering in two dimensions between particles of the same mass is elastic, namely the momenta of the outgoing particles are a permutation of those of the incoming ones.
In particular, for distinguishable particles (such as when scattering gluons of different helicity) two configurations are allowed: forward scattering $p'_1=p_1$, $p'_2 = p_2$, and backward scattering $p'_1=p_2$, $p'_2 = p_1$.
A feature of the gluon-gluon S-matrix is that it is reflectionless, that is backward scattering is absent.
This is a peculiarity that we want to test perturbatively.
Anyway, this means that the scattering of gluons is completely specified by the two scattering factors $S_{gg}$ and $S_{gg^*}$.
Moreover, thanks to 2-dimensional kinematics these are functions of only two independent parameters, that we choose to be the rapidities $\theta_1$ and $\theta_2$.
Imposing momentum conservation explicitly produces a Jacobian from the $\delta$ functions
\begin{equation}\label{eq:Jacobian}
J^{-1} = 4\, e_1(\ppp_1) e_2(\ppp_2) \left( \frac{{\rm d}\, e_1(\ppp_1)}{{\rm d}\, \ppp_1}-\frac{{\rm d}\, e_2(\ppp_2)}{{\rm d}\, \ppp_2}\right)
\end{equation}
which we have to multiply the result of the Feynman diagrams by.
As a final remark, due to the normalization of the action, we introduce an additional factor $N_{x}=\sqrt{2/T}$ for each external gluon, which balances the dependence on the coupling constant correctly.

Hence we finally have the strong coupling expansion of the scattering amplitude for gauge excitations
\begin{equation}
S(\theta_1,\theta_2) = 1 + \frac{i}{g}\, S^{(0)}(\theta_1,\theta_2) + \frac{i}{g^2}\, S^{(1)}(\theta_1,\theta_2) + {\cal O}(g^{-3})
\end{equation}
where we have defined the coupling $g\equiv \frac{\sqrt{\lambda}}{4\pi}$.
For the one-loop term we write
\begin{equation}\label{eq:S}
S^{(1)}(\theta_1,\theta_2) = \frac{J(\theta_1,\theta_2)}{4\pi i}\, {\cal A}^{(1)}(\theta_1,\theta_2)
\end{equation}
where we have factorized the Jacobian, a factor of $i$, and a convenient common factor $4\pi$, emerging from all one-loop integrals.

\section{Tree-level amplitudes}\label{sec:tree}

In this section we briefly review the tree-level computation of gluon amplitudes \cite{Bianchi:2015iza} and provide their result. This is also used to derive a contribution appearing in the one-loop correction.

For scattering of two gluon excitations of the same helicity we evaluate the diagrams of Figure \ref{fig:treexx}.
\FIGURE{
\centering
\includegraphics[width=0.7\textwidth]{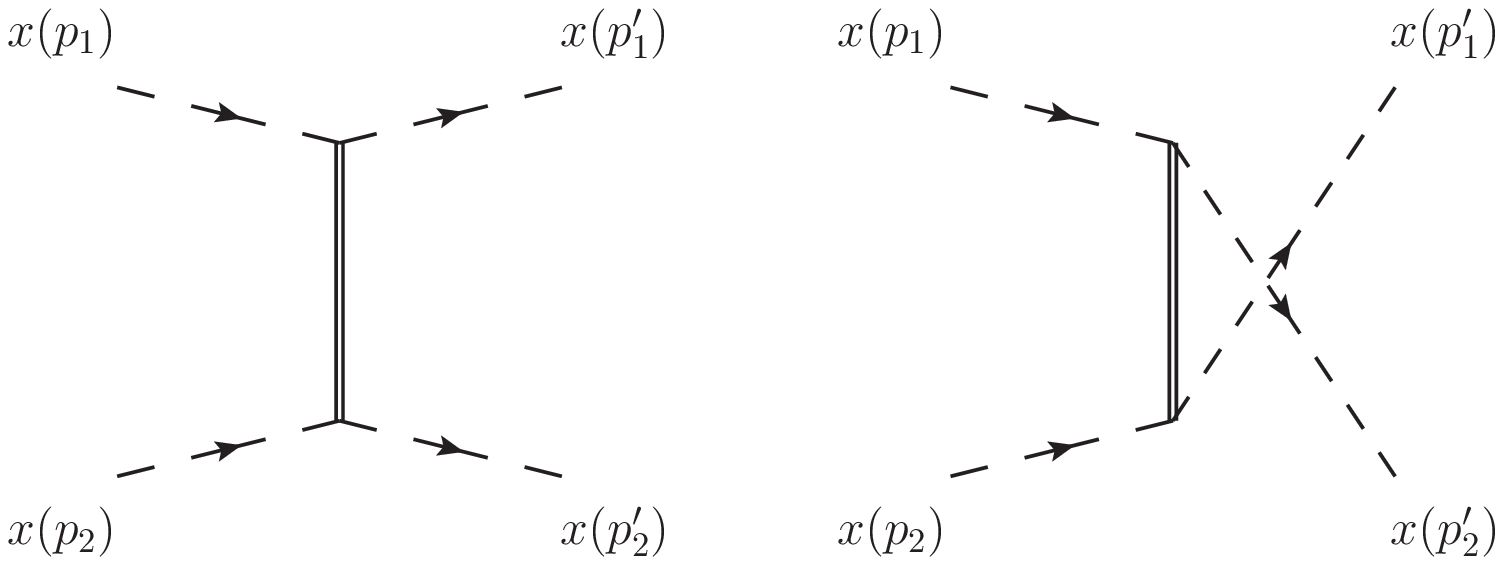}
\caption{Tree-level diagrams for $xx\to xx$ scattering. The exchanged particle is the $\phi$ scalar.}
\label{fig:treexx}
}
The result of the computation of the graphs reads 
\begin{equation}\label{eq:treediaggg}
{\cal A}_{gg}^{(0)}(p_1,p_2) = 8g\, \left(\ppp_1^2+1\right) \left(\ppp_2^2+1\right) \left( \frac{1}{4} + \frac{1}{(p_1-p_2)^2+4} \right) + {\cal O}(g^0)
\end{equation}
and the final S-matrix element in terms of hyperbolic rapidities is
\begin{equation}\label{eq:treeamplitude2}
S_{gg}(\th_1,\th_2) = 1 + \frac{i}{g}\, \frac{\cosh \left(\theta _1-\theta _2\right)+1}{2 \left(\tanh 2 \theta _1 - \tanh 2 \theta _2\right)} + {\cal O}(g^{-2})
\end{equation}

With opposite helicities we have the diagrams in Figure \ref{fig:treexxb}. 
\FIGURE{
\centering
\includegraphics[width=0.7\textwidth]{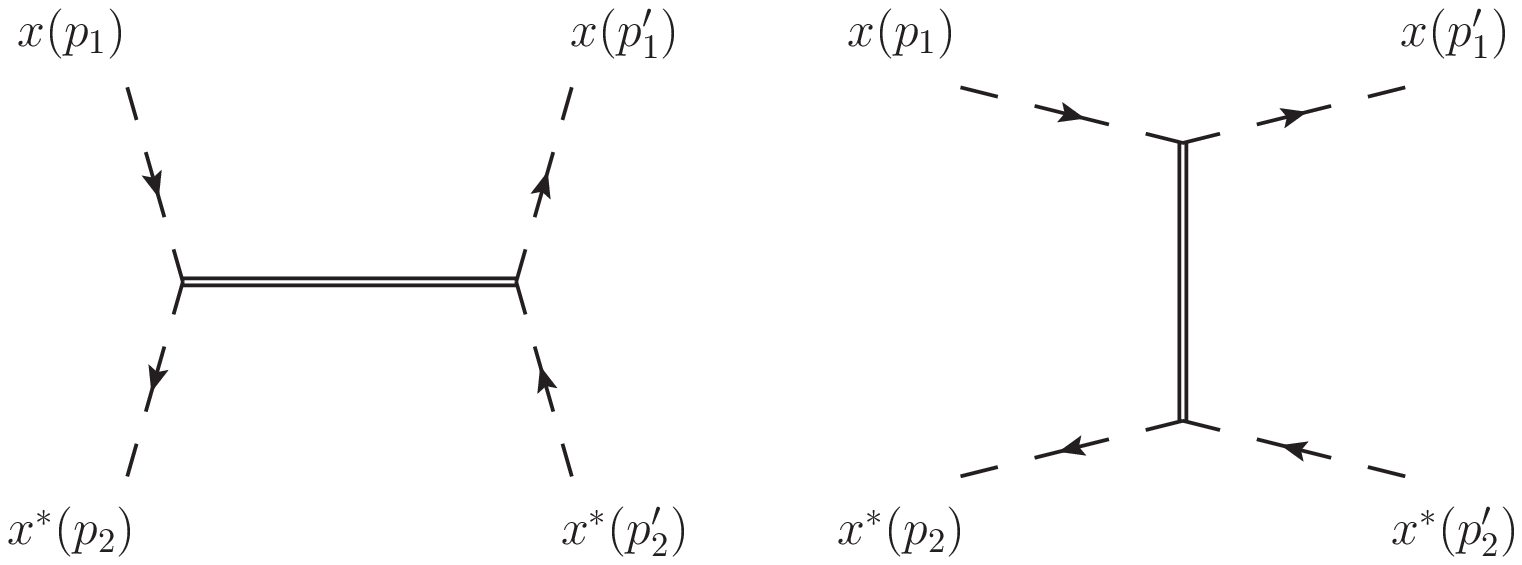}
\caption{Tree-level diagrams for $xx^*\to xx^*$ scattering. The exchanged particle is the $\phi$ scalar.}
\label{fig:treexxb}
}
For the forward solution to the kinematic constraints the Feynman diagrams give
\begin{equation}\label{eq:treexxbforw}
{\cal A}_{gg^*}^{(0)}(p_1,p_2) = 8g\, \left(\ppp_1^2+1\right) \left(\ppp_2^2+1\right) \left( \frac{1}{4} + \frac{1}{(p_1+p_2)^2+4} \right) + {\cal O}(g^0)
\end{equation}
which yields the final amplitude
\begin{equation}\label{eq:treexxb}
S_{gg^*}(\th_1,\th_2) = 1+\frac{i}{g}\, \frac{\cosh \left(\theta _1-\theta _2\right)-1}{2 \left(\tanh 2 \theta _1 - \tanh 2 \theta _2\right)} + {\cal O}(g^{-2})
\end{equation}
In the backward scattering kinematics the amplitude vanishes due to a cancellation between the two exchange channels.
Explicitly the diagrams read
\begin{equation}\label{eq:treexxbback}
\overleftarrow{{\cal A}}_{gg^*}^{(0)}(p_1,p_2) = 8g\, \left(\ppp_1^2+1\right) \left(\ppp_2^2+1\right) \left(\frac{1}{(p_1+p_2)^2+4} + \frac{1}{(p_1-p_2)^2+4}\right) = 0
\end{equation}
where the last equality follows from the identity
\begin{equation}\label{eq:kinid}
(p_1+p_2)^2 + 4 = -(p_1-p_2)^2-4
\end{equation}
which holds for mass $\sqrt{2}$ particles.

We point out that these perturbative results are in agreement with the integrability predictions \eqref{eq:predictionsametree} and \eqref{eq:predictionoppositetree}.

\section{Integral reduction(s)}\label{sec:integral}

\subsection{Reduction to lower order topologies}

In two space-time dimensions it is possible to reduce all higher-point one-loop integrals to bubbles and tadpoles.
This gives more compact expressions when casting results in terms of a basis of integrals and provides an easier check against the constraints imposed by unitarity.
The reduction to bubble integrals can be performed for instance via the van Neerven-Vermaseren procedure \cite{vanNeerven:1983vr}.
For the diagrams at hand we have ascertained that bosonic diagrams can be expressed in terms of the following basis of integrals
\begin{align}\label{eq:basisbos}
& \left\{\, {\rm I}[2,2;s],\, {\rm I}[2,2;u],\, {\rm I}[2,2;0],\, {\rm I}[4,4;s],\, {\rm I}[4,4;u],\, {\rm I}[4,4;0],\, {\rm I}[2,4; -2],\, {\rm I}[2],\, {\rm I}[4] \,\right\}
\end{align}
whereas fermionic diagrams evaluate to a rational combination of the scalar bubbles and tadpoles
\begin{equation}\label{eq:basisfer}
\left\{\, {\rm I}[1,1;s],\, {\rm I}[1,1;u],\, {\rm I}[1,1;0],\, {\rm I}[1,1;-2],\, {\rm I}[1] \,\right\}
\end{equation}
The Mandelstan variables are defined as
\begin{equation}
 s=(p_1+p_2)^2 \qquad u=(p_1-p_2)^2
\end{equation}
and, on-shell, they are related by momentum conservation $s+u=-8$, leaving one independent degree of freedom.
In \eqref{eq:basisbos} and \eqref{eq:basisfer} we have used the notation
\begin{align}
{\rm I}[m_1^2,m_2^2;p^2] &\equiv \int \frac{d^2l}{(2\pi)^2}\, \frac{1}{\left[ l^2 + m_1^2 \right]\left[ (l+p)^2 + m_2^2 \right]}\\
{\rm I}[m^2] &\equiv \int \frac{d^2l}{(2\pi)^2}\, \frac{1}{l^2 + m^2}
\end{align}
The former integral is finite (provided masses are non-vanishing) whereas the latter is UV divergent.
For bubbles the following result is handy ($p$ here, as usual, is to be interpreted as the norm of the two-momentum and not as a vector)
\begin{equation}\label{eq:bubbleintegral}
{\rm I}[m_1^2,m_2^2;p^2] = \frac{\log \frac{p^2+m_1^2+m_2^2+\sqrt{(p^2+m_1^2+m_2^2)^2-4\, m_1^2\, m_2^2}}{p^2+m_1^2+m_2^2-\sqrt{(p^2+m_1^2+m_2^2)^2-4\, m_1^2\, m_2^2}}}{4\,\pi\, \sqrt{(p^2+m_1^2+m_2^2)^2-4\, m_1^2\, m_2^2}}
\end{equation}
from which we have in particular
\begin{equation}\label{eq:somebubbles}
{\rm I}[2,4;-2] = \frac{1}{32}\qquad\qquad
{\rm I}[1,1;-2] = \frac{1}{8}\qquad\qquad
{\rm I}[m^2,m^2;0] = \frac{1}{4\pi\,m^2}
\end{equation}
Notice that the first two integrals in \eqref{eq:somebubbles} are bubble with ingoing momentum $p_1$ or $p_2$ set to its on-shell value $p_i^2=-2$. 
Tadpoles can be computed in dimensional regularization and give
\begin{equation}
{\rm I}[m^2] = \frac{1}{4\pi}\left(\frac{1}{\epsilon} - \log m^2\right) + {\cal O}(\epsilon)
\end{equation}
with a suitable normalization discarding extra unwanted constants that cancel out in the final result.

The evaluation of the latter integrals shows the possible numbers which can appear in the one-loop amplitude.
Integrals with an external momentum in the loop are responsible for terms proportional to $\pi$. Tadpoles (and what remains from finite combinations thereof) generate $\log 2$ terms.
Finally integrals with vanishing momentum (which are nothing but tadpoles with a squared propagator), produce algebraic numbers of lower transcendentality.
Such kind of terms can be also produced by evanescent terms multiplying divergent tadpoles, which are ubiquitous in dimensional regularization. Consequently, we anticipate that this kind of terms are scheme dependent.
On the contrary, the transcendental numbers described above are scheme independent and possess a physical meaning.
We elaborate more on scheme dependence issues in Section \ref{sec:tensor}.

In order to reduce all integrals to the basis above, two steps are needed. First, tensor integrals have to be reduced to scalar integrals. We provide more details on this step in the following section. Second, scalar integrals with higher number of propagators are reduced to bubbles and tadpoles. We spell out how this is achieved for the relevant integrals appearing in the computation.
These can be classified in terms of the number of propagators and are box, triangle, bubble and tadpole integrals.

\paragraph{Triangles}

The scalar triangle integrals appearing in the computation are reduced as follows
\begin{align}
& \int \frac{d^2l}{(2\pi)^2} \frac{1}{(l^2+2)[(l+p_1)^2+4][(l+p_2)^2+4]} = \frac14\, {\rm I}[4,4;u]\\
& \int \frac{d^2l}{(2\pi)^2} \frac{1}{(l^2+1)[(l+p_1)^2+2][(l+p_2)^2+2]} = \frac{(8+s){\rm I}[2,2;u] - 8\,{\rm I}[2,4;-2]}{8(s+4)} \\
& \int \frac{d^2l}{(2\pi)^2} \frac{1}{(l^2+1)[(l+p_1)^2+1][(l+p_2)^2+1]} = -\frac{2}{s+4}\, {\rm I}[1,1;-2] + \frac12\, {\rm I}[1,1;u]
\end{align}
All other triangle integrals which are relevant for the one-loop computation can be obtained by either replacing, e.g., $p_2\to -p_2$ (and consequently $s\to u$) or in the limit $p_2\to p_1$.
The relations above have been derived via the van Neerven-Vermaseren formalism and checked analytically and numerically solving the relevant integrals.

\paragraph{Boxes}

The scalar box integrals appearing in the Feynman diagrams for gluon-gluon scattering are in a special kinematic configuration since scattering in two dimensions is elastic.
In particular the first class of them is actually a triangle topology with a squared propagator.
Such integrals are
\begin{align}
&\int \frac{d^2l}{(2\pi)^2} \frac{1}{(l^2+2)^2[(l+p_1)^2+4][(l+p_2)^2+4]} = \nonumber\\
&~~~~= \frac{(s-4) {\rm I}[4,4;u] + 4 {\rm I}[2,4;-2]}{8 s}\\
&\int \frac{d^2l}{(2\pi)^2} \frac{1}{(l^2+4)^2[(l+p_1)^2+2][(l+p_2)^2+2]} = \nonumber\\
&~~~~= \frac{\left(s^2+8 s+32\right) {\rm I}[2,2;u] - 32\, {\rm I}[2,4;-2] -8\, (s+4) {\rm I}[4,4;0])}{32 (s+4)^2}\\
&\int \frac{d^2l}{(2\pi)^2} \frac{1}{(l^2+1)^2[(l+p_1)^2+1][(l+p_2)^2+1]} = \nonumber\\
&~~~~= \frac{(s+2) {\rm I}[1,1;u] - 4\, {\rm I}[1,1;-2] - 2\, {\rm I}[1,1;0]}{2 (s+4)}
\end{align}
Again, other slightly different integrals appear in the computation, which can be dealt with by swapping the sign of one of the external momenta.
Then there are proper box topologies, but with vanishing $t$-channel.
These read
\begin{align}
&\int \frac{d^2l}{(2\pi)^2} \frac{1}{(l^2+2)[(l+p_1)^2+4][(l+p_1+p_2)^2+2][(l+p_2)^2+4]} =  \nonumber\\
&~~~~= -\frac{s {\rm I}[2,2;s] - 2 (s+4) {\rm I}[4,4;u] + 8\, {\rm I}[2,4;-2]}{4 s (s+4)}\\
&\int \frac{d^2l}{(2\pi)^2} \frac{1}{(l^2+1)[(l+p_1)^2+1][(l+p_1+p_2)^2+1][(l+p_2)^2+1]} =  \nonumber\\
&~~~~=\frac{(s+4) ({\rm I}[1,1;u] - {\rm I}[1,1;s]) - 8\, {\rm I}[1,1;-2]}{(s+4)^2}
\end{align}

\subsection{Tensor reduction}\label{sec:tensor}

Due to the derivative interactions in the Lagrangian \eqref{eq:lagrangian} and fermion propagators, the Feynman diagrams produce tensor integrals.
Power counting shows that the maximum number of momenta in the numerator is four.
When this occurs for triangles or bubbles the integrals are UV divergent.

Tensor integrals have to be reduced to scalar ones.
There are different approaches which can be followed to perform this reduction. They differ by scheme dependent terms and one should try to find one which reproduces the integrability result, at least at this perturbative order.

In the first place we use the most traditional approach, namely Passarino-Veltman reduction \cite{Passarino:1978jh}.
We note that for the integrals at hand PV determinants are singular in strictly two dimensions. As a result we have to deal with PV reduction in $d=2-2\epsilon$ dimensions.
At the end of the reduction process one encounters several integrals with inverse propagators, which can be reduced to master integrals. We have performed this step both via the automated algorithm based on IBP identities \texttt{FIRE} \cite{Smirnov:2008iw} and by hand.
The final master integrals can then be further reduced to bubbles and tadpoles as explained above.
Since the reduction is performed in $d$ dimensions, there are ubiquitous factors of the regularization parameter $\epsilon$ in the expressions, which can then be expanded in series. Even in finite integrals such terms may hit tadpole integrals and produce rational pieces (proportional to algebraic numbers).
Consistently, keeping track of all such factors, the final result in terms of bubble integrals coincides with the original (finite) tensor integral as evaluated directly by, e.g., Feynman parameters. As a check, we have verified that this is indeed the case, for all finite tensor integrals, integrating numerically in the regions where the integrals converge.

We could also have followed a different prescription where the PV reduction is performed in $d$ dimensions, but we then take $d=2$ directly, thus discarding the evanescent terms described above. This is not a consistent procedure of regularization in the sense that when applied to finite integrals it does not yield the correct result for it.
Nevertheless it differs from the previous method only by scheme dependent terms and thus could be an acceptable prescription to deal with numerators.

As we discussed in the Introduction, dimensional regularization might not be an ideal scheme for these kinds of two-dimensional integrable models. 
From a {\it a priori} point of view this can be understood from the fact that a crucial symmetry for the classical integrability of the model, i.e. $\kappa$-symmetry, is chiral (and has a self-dual parameter) and therefore is defined in strictly two dimensions. This is related to the fact that the string sigma model action contains a parity odd Wess-Zumino term, proportional to a two-dimensional Levi-Civita symbol. Although we do not discuss them here, let us mention that possible recipes to analytically continue the two-dimensional Levi-Civita symbol are present in the literature (see, e.g., \cite{deWit:1993qv}). In addition, it can be verified that dimensional regularization does not preserve some two-dimensional algebraic identities at the level of the numerator of the integrands, which one might want to enforce.

On the other hand, from a {\it a posteriori} point of view, we observe that the application of dimensional regularization to the computation of two-point functions and S-matrices pollutes the result with scheme dependent terms which are not present in the integrability prediction. 

An alternative treatment of numerators has been proposed in \cite{Roiban:2014cia}.
There the bottomline is to reduce numerators by using two algebraic identities valid in two dimensions.
In the case of \cite{Roiban:2014cia} this was applied to bubble and tadpole integrals with a single mass.
Nevertheless there is no obstruction in extending this procedure to integrals with a higher number of propagators and different masses.
In \cite{Roiban:2014cia} the authors use light-cone momenta to which we can always switch,
reexpressing the numerators with the momentum components $l_0$ and $l_1$ by light-cone momenta $l_{\pm} \equiv l_0 \pm i\, l_1$.
The resulting integrals can be classified by the propagators and the number of momenta $(n_{l_+},n_{l_-})$ appearing in the numerator.
Then it is immediate to write algebraic relations, such as
\begin{align}\label{eq:id1}
l_+ l_- = [l^2 + m_1^2] - m_1^2 = [(l+p)^2+m_2^2] - l_+ p_- - l_- p_+ - p^2 - m_2^2
\end{align}
and
\begin{equation}\label{eq:id2}
l_+ p_- + l_- p_+ = \left[(l+p)^2 + m_2^2\right] - \left[ l^2 + m_1^2 \right] - \left( p^2 + m_2^2 - m_1^2 \right)
\end{equation}
Imposing such relations on the integrand of the integrals appearing in our computation, one can iteratively reduce the powers of numerators and arrive at a minimal set of integrals.
More explicitly, whenever we have an integral with indices $(n+k,n)$ and $n\geq 1$, we can apply the identity \eqref{eq:id1} and express it in terms of numerators with indices $(n+k-1,n-1)$, $(n+k,n-1)$ and $(n+k-1,n-1)$.

While performing this reduction, integrals of lower topology are generated because of the inverse propagators on the right-hand-side of \eqref{eq:id1}.
The reiterated application of \eqref{eq:id1} breaks whenever an integral with an index 0 is reached or the lowest order topology, a tadpole, is generated.
The latter can be assumed to be irreducible by algebraic relations and can finally be evaluated by dimensional regularization. The indices of such tadpoles vary according to the powers and propagators of the original integrals.
Hence, we can assume that all integrals resulting from the first step should either be irreducible tadpoles or be in the form $(k,0)$ or $(0,k)$.
On these we can further apply \eqref{eq:id2}.
For integrals with indices $(k,0)$ ($k>1$) we interpret one of the indices as the $l_+$ appearing on the l.h.s. of that identity and after imposing \eqref{eq:id2} we can express it in terms of a bunch of integrals with indices $(k-1,1)$ and $(k-1,0)$. The first class falls into the category which is reducible by \eqref{eq:id1}.

Finally, iteratively applying this procedure we arrive at a basis of master integrals with one power of momentum at most.
This procedure can be implemented algorithmically and provides a very efficient reduction.
In particular, starting from boxes and triangles with at most four powers of momentum in the numerator, we reduce them to boxes and triangles with one power of momentum at most in the numerator.
These integrals are finite and can then be straightforwardly evaluated, for instance with the techniques described above.

We have applied this procedure to the relevant integrals and have compared the different reduction methods.
Importantly, we find that for all integrals the coefficients of the tadpoles and the bubbles with invariants $s$, $u$ and $-2$ coincide in all schemes.
This means that the part of the amplitude proportional to these is indeed scheme independent as expected.
We recall that these integrals account for the maximally transcendental part of the amplitude, with potential logarithms of the kinematic invariants and terms proportional to the transcendental constants $\pi$ and $\log 2$.
The different reduction procedures differ for bubbles with 0 momentum invariant and rational terms  of lower transcendentality.
In particular we observe empirically that for all box (even the degenerate ones) and triangle topologies the RSTW scheme produces the same results as from the PV reduction in two dimensions.

\section{One-loop same helicity scattering}\label{sec:samehelicity}

\subsection{Diagrams}

In this section we list and compute the Feynman diagrams contributing to the same helicity amplitude at one loop.
We divide them into topologies, namely boxes, triangles, bubbles and tadpoles.

\paragraph{Boxes}
The box diagrams are depicted in Figure \ref{fig:boxes}.
\FIGURE[h]{
\centering
\includegraphics[width=1.\textwidth]{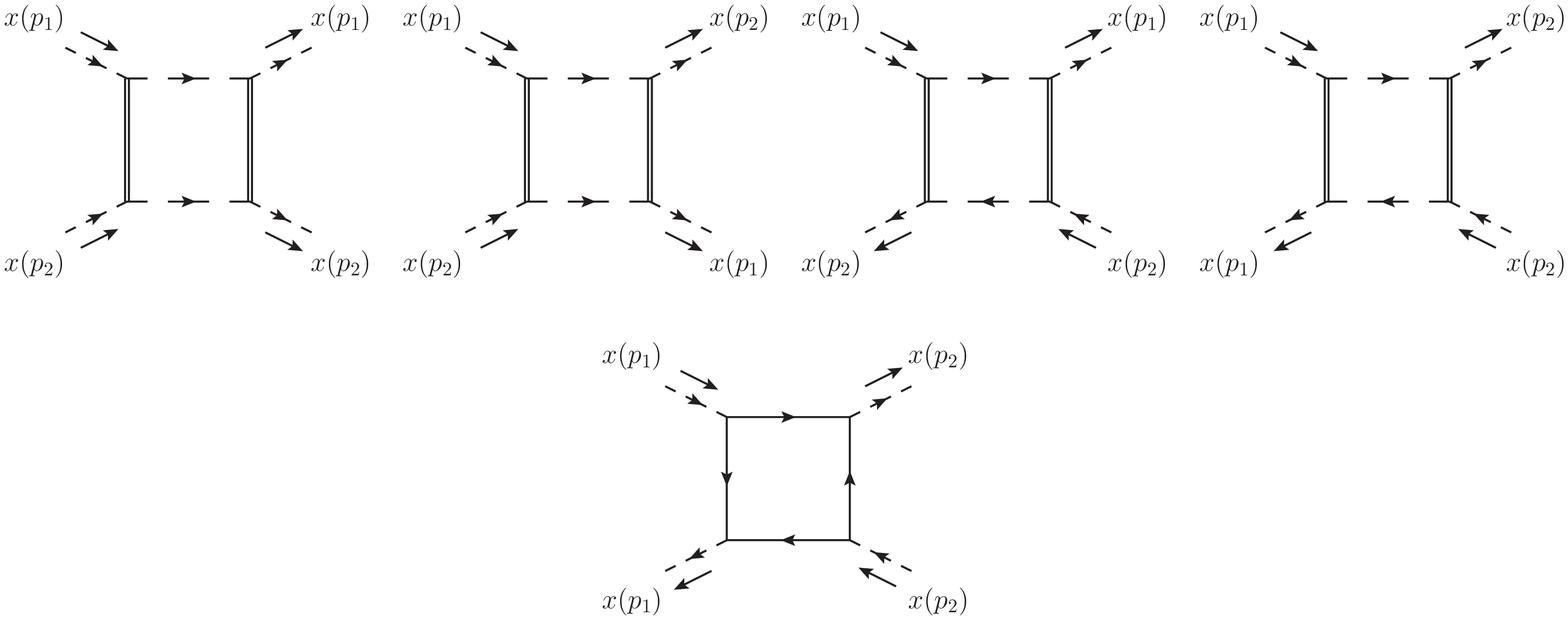}
\caption{Box diagrams for scattering of two gluons of the same helicity. We use the notation of \eqref{eq:props} for particle propagators. Arrows on the propagators indicate how the charge of the particles flows. Whenever ambiguous we put an additional arrow stemming for incoming/outgoing momentum.}
\label{fig:boxes}
}
There are four possible contractions of bosonic diagrams which evaluate\footnote{Here and in the following we denote the loop momentum $l$ as $l=(l_0,l_1)$. Hopefully this will not generate any confusion with the indices $1$ and $2$ associated to the ingoing momenta $p_1=(e_1,\ppp_1)$ and $p_2=(e_2,\ppp_2)$.}
\begin{align}\label{eq:bosonbox}
{\rm Box}_b^{gg} &= 64(\ppp_1^2+1)(\ppp_2^2+1) \bigg\{
\int \frac{d^2l}{(2\pi)^2} \frac{[(l_1-\ppp_1)^2+1][(l_1+\ppp_2)^2+1]}{(l^2+4)^2[(l-p_1)^2+2][(l+p_2)^2+2]} + \nonumber\\&
+ \frac{[(l_1+\ppp_1)^2+1][(l_1+\ppp_2)^2+1]}{(l^2+4)^2[(l+p_1)^2+2][(l+p_2)^2+2]} + \nonumber\\&
+ \frac{(l_1^2+1)[(l_1+\ppp_1+\ppp_2)^2+1]}{(l^2+2)[(l+p_1)^2+4][(l+p_1+p_2)^2+2][(l+p_2)^2+4]} + \nonumber\\&
+ \frac{(l_1^2+1)^2}{(l^2+2)^2[(l+p_1)^2+4][(l+p_2)^2+4]} \bigg\}
\end{align}
and one fermionic box reading
\begin{equation}
{\rm Box}_f^{gg} = -64 (\ppp_1^2+1)(\ppp_2^2+1)  \int \frac{d^2l}{(2\pi)^2} \frac{l_0^2 (l_0+e_1) (l_0+e_2)}{[l^2+1]^2[(l+p_1)^2+1][(l+p_2)^2+1]}
\end{equation}
All integrals are finite.

\FIGURE{
\centering
\includegraphics[width=1.\textwidth]{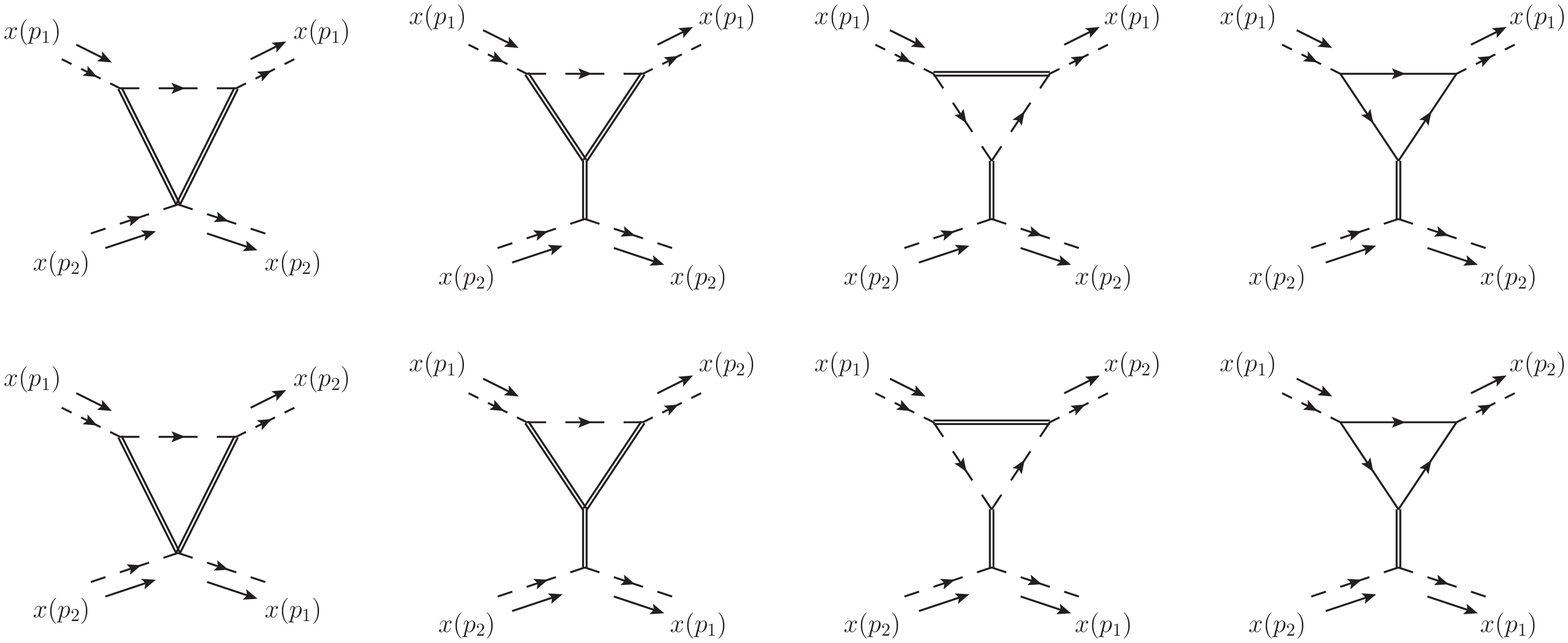}
\caption{Triangle diagrams for scattering of two gluons of the same helicity.}
\label{fig:triangles}
}
\paragraph{Triangles}
Triangle diagrams are shown in Figure \ref{fig:triangles} in a particular configuration.
A permutation of the incoming momenta gives the other diagrams and is already included in the following results.
The first bosonic triangle reads
\begin{align}
{\rm Tri}_1^{gg} &= - 128(\ppp_1^2+1)(\ppp_2^2+1)  \int \frac{d^2l}{(2\pi)^2} \frac{(l_1^2+1)}{(l^2+2)[(l+p_1)^2+4][(l+p_2)^2+4]} + \nonumber\\&
- 64(\ppp_1^2+1)(\ppp_2^2+1)  \int \frac{d^2l}{(2\pi)^2} \frac{(l_1^2+1)}{(l^2+2)[(l+p_1)^2+4]^2} + \nonumber\\&
- 64(\ppp_1^2+1)(\ppp_2^2+1)  \int \frac{d^2l}{(2\pi)^2} \frac{(l_1^2+1)}{(l^2+2)[(l+p_2)^2+4]^2}
\end{align}
and is finite.
The second triangle evaluates
\begin{align}
{\rm Tri}_2^{gg} &= - \frac{64(\ppp_1^2+1)(\ppp_2^2+1)}{(p_1-p_2)^2 + 4}  \int \frac{d^2l}{(2\pi)^2} \frac{(l_1^2+1)[l_0^2+l_0(e_1+e_2)+e_1^2+e_2^2-e_1e_2 - (t\leftrightarrow s)]}{(l^2+2)[(l+p_1)^2+4][(l+p_2)^2+4]} + \nonumber\\&
- \frac{32(\ppp_1^2+1)(\ppp_2^2+1)}{4}  \int \frac{d^2l}{(2\pi)^2} \frac{(l_1^2+1)[(l_0+e_1)^2-(l_1+\ppp_1)^2]}{(l^2+2)[(l+p_1)^2+4]^2} + \nonumber\\&
- \frac{32(\ppp_1^2+1)(\ppp_2^2+1)}{4}  \int \frac{d^2l}{(2\pi)^2} \frac{(l_1^2+1)[(l_0+e_2)^2-(l_1+\ppp_2)^2]}{(l^2+2)[(l+p_2)^2+4]^2}
\end{align}
and the integrals with four powers of loop momentum are UV divergent. The antisymmetrization indicated by $(t\leftrightarrow s)$ is between time and space indices.
The last bosonic topology gives
\begin{align}
{\rm Tri}_3^{gg} & = \frac{128(\ppp_1^2+1)(\ppp_2^2+1)}{(p_1-p_2)^2 + 4}  \int \frac{d^2l}{(2\pi)^2} \frac{[(l_1+\ppp_1)^2+1][(l_1+\ppp_2)^2+1]}{(l^2+4)[(l+p_1)^2+2][(l+p_2)^2+2]} + \nonumber\\&
+ \frac{64(\ppp_1^2+1)(\ppp_2^2+1)}{4}  \int \frac{d^2l}{(2\pi)^2} \frac{[(l_1+\ppp_1)^2+1]^2}{(l^2+4)[(l+p_1)^2+2]^2} + \nonumber\\& + \frac{64(\ppp_1^2+1)(\ppp_2^2+1)}{4}  \int \frac{d^2l}{(2\pi)^2} \frac{[(l_1+\ppp_2)^2+1]^2}{(l^2+4)[(l+p_2)^2+2]^2}
\end{align}
and is again divergent.
Finally there is a fermion loop diagram
\begin{align}
{\rm Tri}_4^{gg} &= \frac{128(\ppp_1^2+1)(\ppp_2^2+1)}{(p_1-p_2)^2 + 4} \int \frac{d^2l}{(2\pi)^2} \frac{l_0(l_0+e_1)[(l_1+\ppp_2)^2+1]+l_0(l_0+e_2)[(l_1+\ppp_1)^2+1]}{(l^2+1)[(l+p_1)^2+1][(l+p_2)^2+1]} + \nonumber\\&
+ \frac{128(\ppp_1^2+1)(\ppp_2^2+1)}{4} \int \frac{d^2l}{(2\pi)^2} \frac{l_0(l_0+e_1)[(l_1+\ppp_1)^2+1]}{(l^2+1)[(l+p_1)^2+1]^2} + \nonumber\\& + \frac{128(\ppp_1^2+1)(\ppp_2^2+1)}{4} \int \frac{d^2l}{(2\pi)^2} \frac{l_0(l_0+e_2)[(l_1+\ppp_2)^2+1]}{(l^2+1)[(l+p_2)^2+1]^2}
\end{align}
which is divergent.

\paragraph{Bubbles}
Bubble diagrams are sketched in  Figure \ref{fig:bubbles} for one configuration.
\FIGURE{
\centering
\includegraphics[width=1.\textwidth]{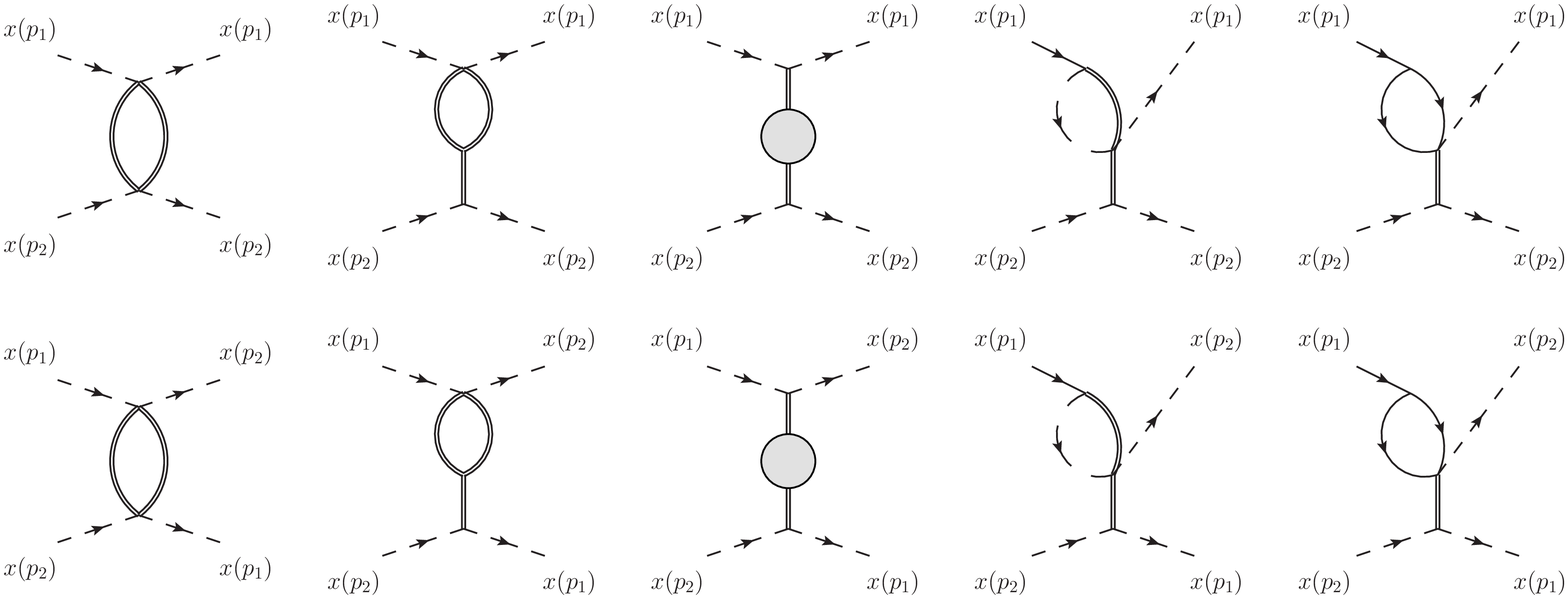}
\caption{Bubble diagrams for scattering of two gluons of the same helicity.}
\label{fig:bubbles}
}
A permutation (affecting all diagrams but the first and the third) has to be considered, which is already included in the results below (and amounts to a factor 2 for the second and 4 for the last two diagrams).
The first bosonic bubble diagram evaluates
\begin{equation}
{\rm Bub}_1^{gg} = 32 (\ppp_1^2+1)(\ppp_2^2+1) \left[ {\rm I}[4,4;0]+{\rm I}[4,4;u] \right]
\end{equation}
and is manifestly finite.
The second reads
\begin{align}\label{bub2}
{\rm Bub}_2^{gg} &= 32 (\ppp_1^2+1)(\ppp_2^2+1) \bigg[ \frac14 \int \frac{d^2l}{(2\pi)^2} \frac{l_0^2-l_1^2}{(l^2+4)^2} + \nonumber\\& 
+ \frac{1}{(p_1-p_2)^2+4} \int \frac{d^2l}{(2\pi)^2} \frac{l_0^2-l_0(e_1-e_2)+(e_1-e_2)^2 - (t \leftrightarrow s)}{(l^2+4)^2[(l+p_2-p_1)^2+4]} \bigg]
\end{align}
and is in principle UV divergent by power counting, though the divergence cancels due to $t$, $s$ antisymmetry.
For the same reason the first integral in \eqref{bub2} vanishes identically, giving
\begin{align}
{\rm Bub}_2^{gg} &= \frac{32 (\ppp_1^2+1)(\ppp_2^2+1)}{(p_1-p_2)^2+4} \int \frac{d^2l}{(2\pi)^2} \frac{l_0^2 + l_0(e_1+e_2) - e_1 e_2 + e_1^2 + e_2^2 - (t \leftrightarrow s)}{[(l+p_1)^2+4][(l+p_2)^2+4]}
\end{align}
The third topology involves the (off-shell) one-loop correction to the heavy scalar propagator:
\begin{equation}
{\rm Bub}_3^{gg} = 4 (\ppp_1^2+1)(\ppp_2^2+1) \left[ \braket{\phi(0)\phi(0)}^{(1)} + \braket{\phi(p_1-p_2)\phi(-p_1+p_2)}^{(1)} \right]
\end{equation}
This contribution is UV finite.
We can use the results of \cite{Giombi:2010bj} for the integrand of the two-point function to construct the diagram.
Therefore we have for the self-energy corrections
\begin{align}
& \langle \phi(P)\phi(-P)\rangle^{(1)} = 4\, \frac{(\ppp_1^2+1)(\ppp_2^2+1)}{(p_1-p_2)^2+4} \int \frac{d^2l}{(2\pi)^2} \bigg[ \\&
s^{\a\b}s^{\g\d}\frac{2l_\a l_\g(l_\b l_\d+2P_\b l_\d-3P_\b P_\d)+4l_\a P_\b P_\g P_\d}{(l^2+4)[(l-(p_1-p_2))^2+4]} + \nonumber\\&
+16 \frac{(l_1^2+1)[(l_1-(\ppp_1-\ppp_2))^2+1]}{(l^2+2)[(l-(p_1-p_2))^2+2]} - 32 \frac{[(l_1-(\ppp_1-\ppp_2))^2+1][l_1^2+1+l_0(e_1-e_2-l_0)]}{(l^2+1)[(l-(p_1-p_2))^2+1]} + \nonumber\\&
-2 \frac{(p_1-p_2)^2}{l^2+4} - 16 \frac{l_1^2+1}{l^2+2} + 32\frac{l_1^2+1}{l^2+1}
+ 4\frac{(e_1-e_2)^2-(\ppp_1-\ppp_2)^2}{l^2+1} + \nonumber\\&
+ 2\frac{(l_0^2-l_1^2)^2}{(l^2+4)^2} + 16 \frac{(l_1^2+1)^2}{(l^2+2)^2} - 32 \frac{[l_1^2+1][l_1^2-l_0^2+1]}{(l^2+1)^2} - 16 \frac{l_1^2+1}{l^2+2} + 32\frac{l_1^2+1}{l^2+1}
\bigg]
\end{align}
where $s={\rm diag}(1,-1)$ and $P=p_1-p_2$. The contribution at vanishing inflowing momentum can be obtained as a limit.
The reduction can be affected by different scheme choices, as the rest of the computation.
The procedure used in \cite{Giombi:2010bj} is similar to the reduction in strictly two dimensions outlined in Section \ref{sec:tensor}.
We have explicitly verified (by redoing the reduction of \cite{Giombi:2010bj}) that indeed the reduction via dimensional regularization differs from the latter, but only for lower transcendentality rational  terms arising from $\epsilon$-dependent constants multiplying UV divergent tadpoles, as expected.

The last two bubbles have the same topology but differ for the nature of the particles flowing in the loop.
The bosonic diagram reads
\begin{align}
{\rm Bub}_4^{gg} & = -128 (\ppp_1^2+1)(\ppp_2^2+1) \left[ \frac{1}{(p_1-p_2)^2+4} + \frac14 \right] \bigg[ \int \frac{d^2l}{(2\pi)^2} \frac{l_1^2+1}{(l^2+2)[(l+p_1)^2+4]} + \nonumber\\&
+ \int \frac{d^2l}{(2\pi)^2} \frac{l_1^2+1}{(l^2+2)[(l+p_2)^2+4]} \bigg]
\end{align}
and is UV divergent. 
The fermionic loop evaluates
\begin{align}
{\rm Bub}_5^{gg} &= -96 (\ppp_1^2+1)(\ppp_2^2+1) \left[ \frac{1}{(p_1-p_2)^2+4} + \frac14 \right] \bigg[ \int \frac{d^2l}{(2\pi)^2} \frac{l_0(l_0+e_1)}{(l^2+1)[(l+p_1)^2+1]} + \nonumber\\&
+ \int \frac{d^2l}{(2\pi)^2} \frac{l_0(l_0+e_2)}{(l^2+1)[(l+p_2)^2+1]} \bigg]
\end{align}
and is UV divergent too.

\paragraph{Tadpoles}
Finally there are two tadpole diagram topologies to be considered, which are represented in Figure \ref{fig:tadpole}, up to obvious permutations. These can be readily accounted for in the results below and amount to an extra factor 2.
\FIGURE{
\centering
\includegraphics[width=1.\textwidth]{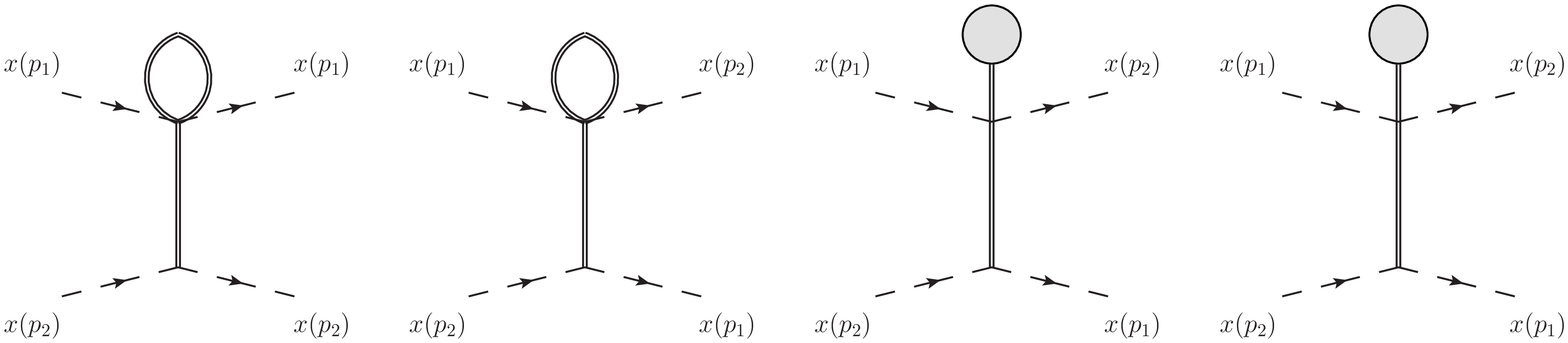}
\caption{Tadpole diagrams for scattering of two gluons of the same helicity.}
\label{fig:tadpole}
}
The first topology consists of a 1PI diagram with a heavy scalar tadpole
\begin{equation}
{\rm Tad}_1^{gg} = 64 (\ppp_1^2+1)(\ppp_2^2+1) \left[ \frac{1}{(p_1-p_2)^2+4} + \frac14 \right] {\rm I}[4]
\end{equation}
The second graph is a reducible topology involving the quantum corrected expectation value of the heavy scalar
\begin{equation}
{\rm Tad}_2^{gg} = -32 (\ppp_1^2+1)(\ppp_2^2+1) \left[ \frac{1}{(p_1-p_2)^2+4} + \frac14 \right] \langle \phi \rangle
\end{equation}
where $\langle \phi \rangle = -2{\rm I}[1]$, so that
\begin{equation}
{\rm Tad}_2^{gg} = 64 (\ppp_1^2+1)(\ppp_2^2+1) \left[ \frac{1}{(p_1-p_2)^2+4} + \frac14 \right] {\rm I}[1]
\end{equation}

\subsection{Expression in terms of bubble integrals}

Bubbles with invariant $s$ are easily seen to emerge only from the box diagrams.
In particular only one such bubble appears, with coefficient
\begin{equation}
{\cal A}_{gg}^{(1)}(p_1,p_2)\Big|_s = 8 \left(\frac{s}{s-u}\right)^2 (\ppp_1^2 + 1)^2 (\ppp_2^2 + 1)^2\, {\rm I}[2,2;s]
\end{equation}
Bubble integrals with invariant $u$ arise from a number of diagrams and the coefficients from each graph and their total sum are collected in Table \ref{tab:samepq} in Appendix \ref{app:details}, where we omit a common factor $(\ppp^2_1+1)(\ppp^2_2+1)=\cosh 2\theta_1 \cosh 2\theta_2$ for brevity. After a non-trivial cancellation among different diagram topologies only ${\rm I}[2,2;u]$ survives and it comes with the same coefficient as ${\rm I}[2,2;s]$
\begin{equation}\label{eq:finalbubbles}
{\cal A}_{gg}^{(1)}(p_1,p_2)\Big|_{\log} = 8 \left(\frac{s}{s -u}\right)^2 (\ppp_1^2 + 1)^2 (\ppp_2^2 + 1)^2 \left( {\rm I}[2,2;s] + {\rm I}[2,2;u] \right)
\end{equation}
This is the part of the amplitude which potentially contains logarithms of the momentum invariants.

The coefficients of bubbles with external momentum $p_1$ or $p_2$ inflowing are spelled out in Table \ref{tab:samep} in Appendix \ref{app:details}.
Summing them up and taking into account the relations \eqref{eq:somebubbles} we find a term proportional to $\pi$ (in units of $4\pi$)
\begin{align}
{\cal A}_{gg}^{(1)}(\theta_1,\theta_2)\Big|_{\pi} &= \frac{\pi}{2} \cosh 2 \theta _1 \cosh 2 \theta _2 \bigg[4 \cosh ^2\left(\theta _1+\theta _2\right)+ \nonumber\\&
+ \frac{\cosh \left(\theta _1-\theta _2\right)+1}{\cosh\left(\theta _1-\theta _2\right)} \left(-\cosh ^2 2 \theta _1 - \cosh ^2 2 \theta _2 + 4 \left(1-\cosh 2 \left(\theta _1+\theta _2\right)\right)\right)\bigg]
\end{align}

Finally there are terms proportional to $\log{2}$ which arise from tadpole integrals.
They come from almost all bosonic diagrams and finally add up to a remarkably simple contribution
\begin{equation}
{\cal A}_{gg}^{(1)}(\theta_1,\theta_2)\Big|_{\log 2} = 2\, \frac{\cosh 2\theta_1\cosh 2\theta_2}{\cosh \left(\theta _1-\theta _2\right)} \cosh ^2\frac{\theta _1-\theta _2}{2} \left(\cosh 4 \theta _1+\cosh 4 \theta _2+8\right) \log{2}
\end{equation}

On top of this there are terms proportional to algebraic numbers that are scheme dependent.
We have computed them in the three different schemes outlined in Section \ref{sec:tensor}.
The coefficients vary significantly according to the scheme, as expected.
In Table \ref{tab:same0} (see Appendix \ref{app:details}) we summarize the coefficients for bubble integrals with momentum 0, given by using the RSTW scheme (see the Introduction).
Using \eqref{eq:somebubbles}, and summing the contributions from all bubbles with different masses, we obtain a remarkably simple expression
\begin{align}\label{eq:samealgebraic}
{\cal A}_{gg}^{(1)}(\theta_1,\theta_2)\Big|_{\rm algebraic} = 2\, \frac{\cosh 2 \theta _1 \cosh 2 \theta _2}{\cosh\left(\theta _1-\theta _2\right)}\, \cosh ^2 \frac{\theta _1-\theta _2}{2}\,  \left(\cosh 4 \theta _1 + \cosh 4 \theta _2 + 2\right)
\end{align}

\subsection{Comments on the result}

We start commenting on finiteness.
First we note that since no coupling with the massless scalars is involved, no IR divergences are generated. An exception is the third bubble topology where actually a $y$ scalar loop appears in the one-loop correction to the two-point function of heavy scalars. Nevertheless, this contribution evaluates to 0 identically and consequently the result is IR finite, as expected.
We remark that although some individual diagrams develop UV divergences, the final sum of all diagrams undergoes a complete cancellation of poles
\begin{equation}
{\cal A}_{gg}^{(1)}\Big|_{\rm UV} = {\cal A}_{gg}^{(1)}\Big|_{\rm IR} = 0
\end{equation}
This was also expected and provides a check on the correctness of the computation.

Next we observe that, quite remarkably, all potential bubble integrals with $s$ and $u$ invariants cancel out of the final result (though a plethora of them appears from tensor reduction of the individual graphs), but those with masses 2.
Furthermore, as a result of \eqref{eq:finalbubbles} and using \eqref{eq:bubbleintegral}, we ascertain that the sum of them is free of logarithms of the momentum invariants. This is because the arguments of the logarithms are opposite and hence the real part vanishes and there survives only a rational imaginary term
\begin{equation}
{\cal A}_{gg}^{(1)}(\theta_1,\theta_2)\Big|_{\log} = i\, \frac{\cosh^2 2\theta_1 \cosh^2 2\theta_2 \cosh^4 \frac{\theta_1-\theta_2}{2}}{\sinh\left(\theta_1-\theta_2\right)\cosh^2\left(\theta_1-\theta_2\right)}
\end{equation}
Since these were the only source of potential logarithms in the computation, the complete amplitude turns out to be rational.
This agrees with the integrability prediction \eqref{eq:predictionsame}.
We can do better and compare the term coming from \eqref{eq:finalbubbles} with the real part of the integrability prediction \eqref{eq:predictionsamereal}.
After including the Jacobian factor and normalization, these pieces are found to coincide.
We recall that in the integrability result such a term arises as the square of the tree-level amplitude by exponentiating the scattering phase.
All these facts are totally consistent with a unitarity based argument, according to which the two-particle cuts in the $s$ and $u$ channels are given by squaring two gluon-gluon tree-level scattering amplitudes.
We leave a more complete description of the construction of this amplitude via unitarity to future studies.

\section{One-loop opposite helicity forward scattering}\label{sec:opphelicityf}

The computation of the one-loop scattering matrix for gluons with opposite helicities in forward kinematics is completely analogous to that for same helicity gauge excitations described in Section \ref{sec:samehelicity}.
In particular, it can be obtained from the latter by a standard crossing transformation
\begin{equation}
 s\leftrightarrow u \qquad p_2\to -p_2
\end{equation}
which allows to interchange the two processes, as expected from crossing symmetry.
Hence we omit a detailed derivation (which we provide in Appendix \ref{app:details}) and simply state the final result.

Similarly to the same helicity scattering, bubbles with invariant $u$ are generated only in the reduction of the bosonic box diagrams, while bubbles with invariant $s$ are ubiquitous and their coefficients are collected in Table \ref{tab:forpq} in Appendix \ref{app:details}.
The final result for bubble integrals with invariant $s$ and $u$ is completely analogous to \eqref{eq:finalbubbles}
\begin{equation}\label{eq:finalbubblesfor}
\overrightarrow{\cal A}_{gg^*}^{(1)}(p_1,p_2)\Big|_{\log} = 8 \left(\frac{u}{s-u}\right)^2 (\ppp_1^2 + 1)^2 (\ppp_2^2 + 1)^2 \left( {\rm I}[2,2;s] + {\rm I}[2,2;u] \right)
\end{equation}
and, as in that case, it has a simple interpretation in terms of unitarity cuts.

The coefficients of the integrals with an external on-shell momentum are collected in Table \ref{tab:forp} in Appendix \ref{app:details}.
Combining the contributions of these integrals we find a term proportional to $\pi$ reading
\begin{align}
\overrightarrow{\cal A}_{gg^*}^{(1)}(\theta_1,\theta_2)\Big|_{\pi}& = \frac{\pi}{2} \cosh 2 \theta _1 \cosh 2 \theta _2 \bigg[4 \cosh ^2\left(\theta _1+\theta _2\right)+ \nonumber\\&
+ \frac{1-\cosh \left(\theta _1-\theta _2\right)}{\cosh\left(\theta _1-\theta _2\right)} \left(\cosh ^2 2 \theta _1 + \cosh ^2 2 \theta _2 - 4 \left(1-\cosh 2 \left(\theta _1+\theta _2\right)\right)\right)\bigg]
\end{align}
Finally there are terms proportional to $\log{2}$ which arise from tadpole integrals.
They come from almost all bosonic diagrams and finally add up to a remarkably simple contribution
\begin{equation}
\overrightarrow{\cal A}_{gg^*}^{(1)}(\theta_1,\theta_2)\Big|_{\log 2} = 2\, \frac{\cosh 2\theta_1\cosh 2\theta_2}{\cosh \left(\theta _1-\theta _2\right)} \sinh ^2 \frac{\theta _1-\theta _2}{2} \left(\cosh 4 \theta _1+\cosh 4 \theta _2+8\right) \log{2}
\end{equation}
On top of this there are scheme dependent terms.
We collect them in Table \ref{tab:opp0} in Appendix \ref{app:details} within the RSTW scheme.
Combining the contributions from all bubbles with different masses, we find
\begin{align}\label{eq:oppalgebraic}
\overrightarrow{\cal A}_{gg^*}^{(1)}(\theta_1,\theta_2)\Big|_{\rm algebraic} = 2\, \frac{\cosh 2 \theta _1 \cosh 2 \theta _2}{\cosh\left(\theta _1-\theta _2\right)}\, \sinh ^2 \frac{\theta _1-\theta _2}{2}\, \left(\cosh 4 \theta _1 + \cosh 4 \theta _2 + 2\right) 
\end{align}

\section{One-loop opposite helicity backward scattering}\label{sec:opphelicityb}

\subsection{Diagrams}

We list and evaluate the Feynman diagram topologies for backward scattering.
\paragraph{Boxes}
The box diagrams are shown in Figure \ref{fig:boxesB}.
\FIGURE[h]{
\centering
\includegraphics[width=1.\textwidth]{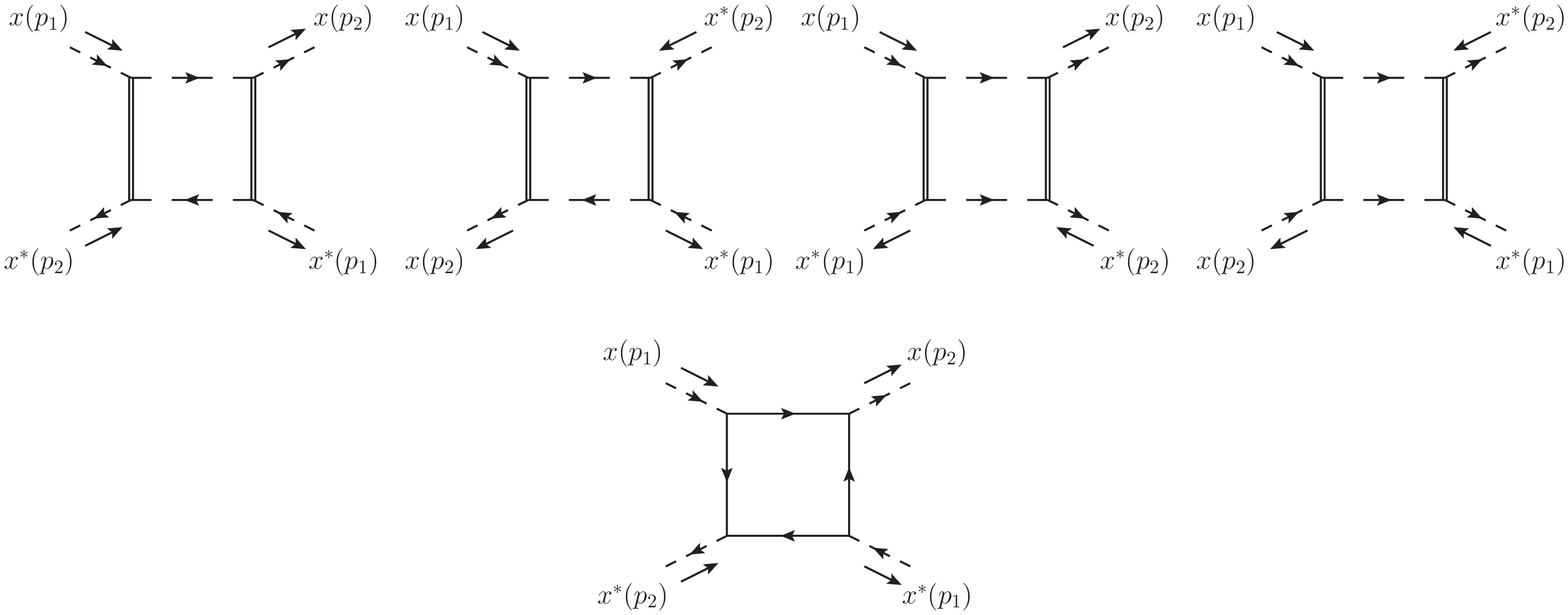}
\caption{Box diagrams in the opposite helicity process and for backward kinematics.}
\label{fig:boxesB}
}
The bosonic part consists of four contributions
\begin{align}\label{eq:bosonboxB}
\overleftarrow{{\rm Box}}_b^{gg^*} &= 64(\ppp_1^2+1)(\ppp_2^2+1) \bigg\{
\frac{(l_1^2+1)[(l_1+\ppp_1+\ppp_2)^2+1]}{(l^2+2)[(l+p_1)^2+4][(l+p_1+p_2)^2+2][(l+p_2)^2+4]} + \nonumber\\&
+ \frac{(l_1^2+1)^2}{(l^2+2)^2[(l+p_1)^2+4][(l+p_2)^2+4]} + \nonumber\\&
+ \frac{(l_1^2+1)[(l_1-\ppp_1+\ppp_2)^2+1]}{(l^2+2)[(l-p_1)^2+4][(l-p_1+p_2)^2+2][(l+p_2)^2+4]} + \nonumber\\&
+ \frac{(l_1^2+1)^2}{(l^2+2)^2[(l-p_1)^2+4][(l+p_2)^2+4]}
\bigg\}
\end{align}
and the fermionic piece reads
\begin{equation}
\overleftarrow{{\rm Box}}_f^{gg^*} = -64 (\ppp_1^2+1)(\ppp_2^2+1)  \int\frac{d^2l}{(2\pi)^2} \frac{l_0 (l_0+e_1) (l_0+e_2) (l_0+e_1+e_2)}{[l^2+1][(l+p_1)^2+1][(l+p_1+p_2)^2+1][(l+p_2)^2+1]}
\end{equation}
This last is a proper fermionic box, at a difference with the previous configurations where the fermionic box always degenerated to triangles with a squared propagator.

\paragraph{Triangles}
The triangle diagrams are collected in Figure \ref{fig:trianglesB} in a particular configuration. A permutation has to be performed, similarly to the forward case. However, in this configuration, it amounts to a factor of 2 for all diagrams.
\FIGURE{
\centering
\includegraphics[width=1.0\textwidth]{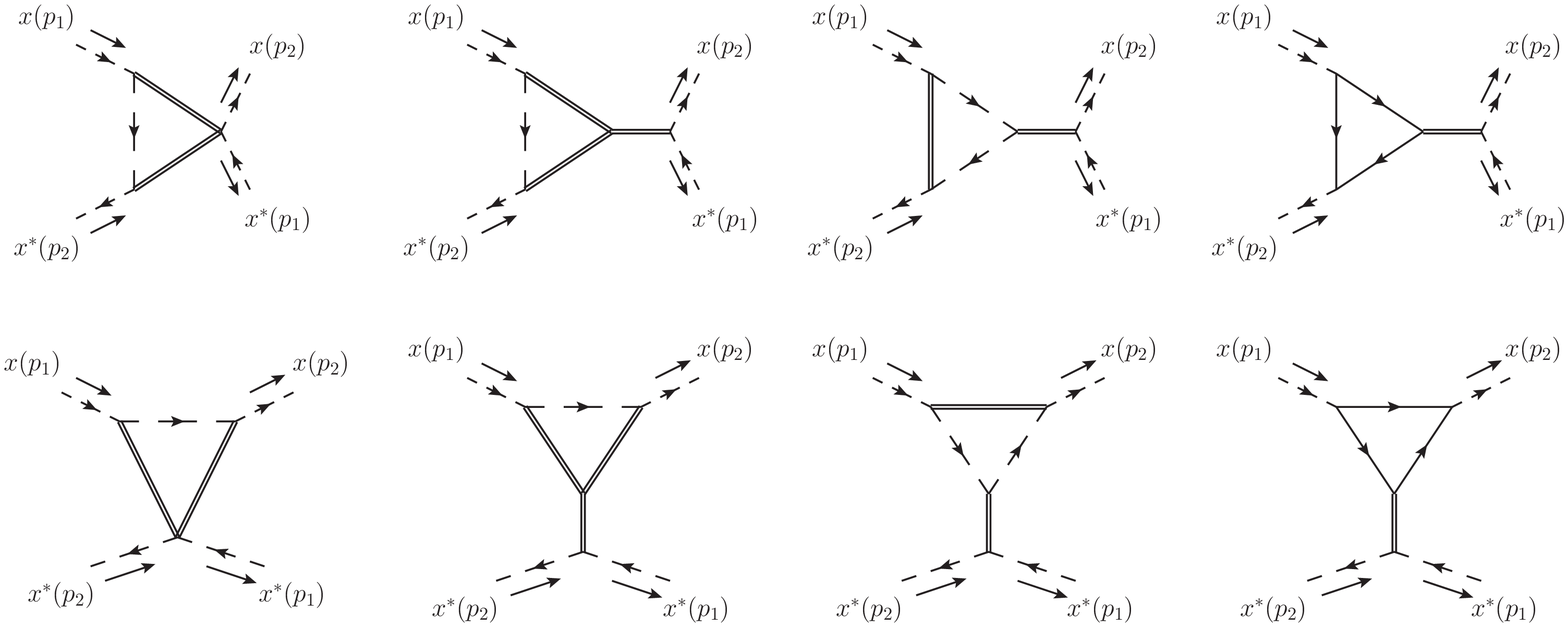}
\caption{Triangle diagrams in the opposite helicity process and for backward kinematics.}
\label{fig:trianglesB}
}
The first topology yields
\begin{align}
\overleftarrow{{\rm Tri}}_1^{gg^*} &= - 128(\ppp_1^2+1)(\ppp_2^2+1)  \int\frac{d^2l}{(2\pi)^2} \frac{(l_1^2+1)}{(l^2+2)[(l+p_1)^2+4][(l+p_2)^2+4]} + \nonumber\\&
- 128(\ppp_1^2+1)(\ppp_2^2+1)  \int\frac{d^2l}{(2\pi)^2} \frac{(l_1^2+1)}{(l^2+2)[(l-p_1)^2+4][(l+p_2)^2+4]}
\end{align}
The second reads
\begin{align}
\overleftarrow{{\rm Tri}}_2^{gg^*} &= - \frac{64(\ppp_1^2+1)(\ppp_2^2+1)}{(p_1-p_2)^2 + 4}  \int\frac{d^2l}{(2\pi)^2} \frac{(l_1^2+1)[l_0^2+l_0(e_1+e_2)+e_1^2+e_2^2-e_1e_2 - (t\leftrightarrow s)]}{(l^2+2)[(l+p_1)^2+4][(l+p_2)^2+4]} + \nonumber\\&
- \frac{64(\ppp_1^2+1)(\ppp_2^2+1)}{(p_1+p_2)^2 + 4}  \int\frac{d^2l}{(2\pi)^2} \frac{(l_1^2+1)[l_0^2+l_0(e_2-e_1)+e_1^2+e_2^2+e_1e_2 - (t\leftrightarrow s)]}{(l^2+2)[(l-p_1)^2+4][(l+p_2)^2+4]}
\end{align}
and the integrals with four powers of loop momentum are UV divergent.
The last bosonic topology gives
\begin{align}
\overleftarrow{{\rm Tri}}_3^{gg^*} &= \frac{128(\ppp_1^2+1)(\ppp_2^2+1)}{(p_1-p_2)^2 + 4}  \int\frac{d^2l}{(2\pi)^2} \frac{[(l_1+\ppp_1)^2+1][(l_1+\ppp_2)^2+1]}{(l^2+4)[(l+p_1)^2+2][(l+p_2)^2+2]} + \nonumber\\&
+ \frac{128(\ppp_1^2+1)(\ppp_2^2+1)}{(p_1+p_2)^2 + 4}  \int\frac{d^2l}{(2\pi)^2} \frac{[(l_1-\ppp_1)^2+1][(l_1+\ppp_2)^2+1]}{(l^2+4)[(l-p_1)^2+2][(l+p_2)^2+2]}
\end{align}
and is again divergent.
Finally there is a fermion loop diagram
\begin{align}
\overleftarrow{{\rm Tri}}_4^{gg^*} &= \frac{128(\ppp_1^2+1)(\ppp_2^2+1)}{(p_1-p_2)^2 + 4}  \int\frac{d^2l}{(2\pi)^2} \frac{l_0(l_0+e_1)[(l_1+\ppp_2)^2+1]+l_0(l_0+e_2)[(l_1+\ppp_1)^2+1]}{(l^2+1)[(l+p_1)^2+1][(l+p_2)^2+1]} + \nonumber\\&
+ \frac{128(\ppp_1^2+1)(\ppp_2^2+1)}{(p_1+p_2)^2 + 4}  \int\frac{d^2l}{(2\pi)^2} \frac{l_0(l_0-e_1)[(l_1+\ppp_2)^2+1]+l_0(l_0+e_2)[(l_1-\ppp_1)^2+1]}{(l^2+1)[(l-p_1)^2+1][(l+p_2)^2+1]}
\end{align}
which is divergent.

\paragraph{Bubbles}
The bubble diagrams of Figure \ref{fig:bubbles} are evaluated similarly to the forward case.
\FIGURE{
\centering
\includegraphics[width=1.\textwidth]{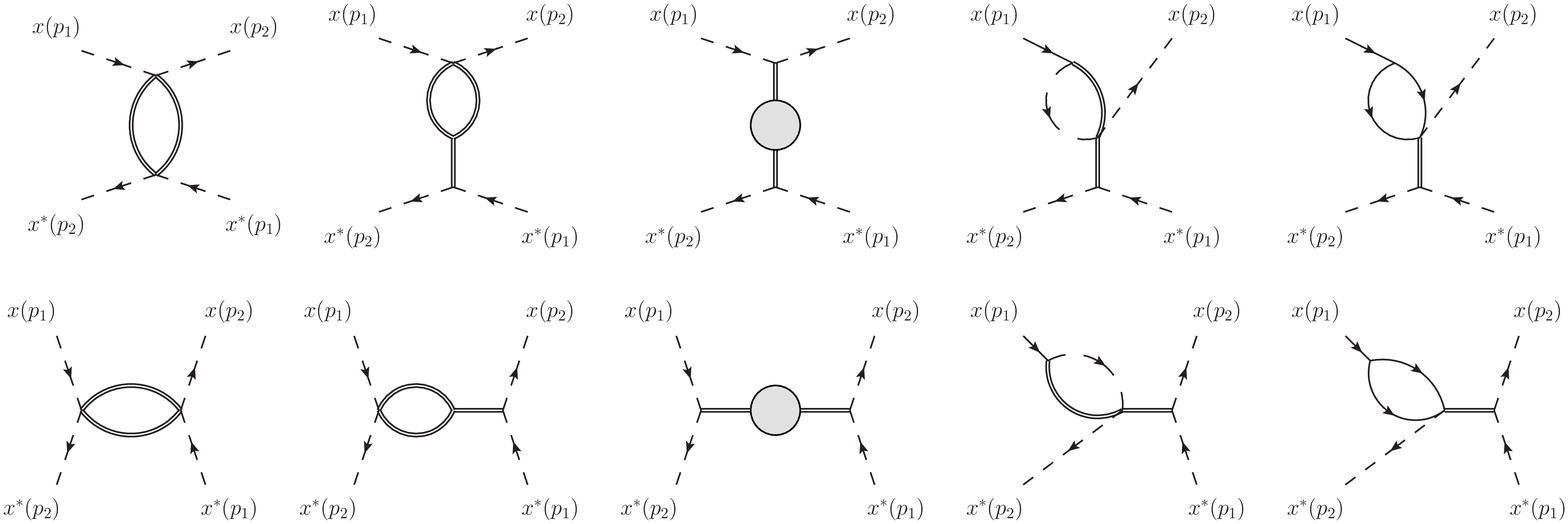}
\caption{Bubble diagrams in the opposite helicity process and for backward kinematics.}
\label{fig:bubblesB}
}

From the first we obtain
\begin{equation}
\overleftarrow{{\rm Bub}}_1^{gg^*} = 32 (\ppp_1^2+1)(\ppp_2^2+1) \left[ {\rm I}[4,4;s]+{\rm I}[4,4;u] \right]
\end{equation}
From the second
\begin{align}
\overleftarrow{{\rm Bub}}_2^{gg^*} &= 32 (\ppp_1^2+1)(\ppp_2^2+1) \bigg[ \frac{1}{(p_1-p_2)^2+4} \int \frac{d^2l}{(2\pi)^2} \frac{l_0^2-l_0(e_1-e_2)+(e_1-e_2)^2 - (t \leftrightarrow s)}{(l^2+4)^2[(l+p_2-p_1)^2+4]} \nonumber\\&
+ \frac{1}{(p_1+p_2)^2+4} \int \frac{d^2l}{(2\pi)^2} \frac{l_0^2 + l_0(e_2-e_1) + e_1 e_2 + e_1^2 + e_2^2 - (t \leftrightarrow s)}{[(l-p_1)^2+4][(l+p_2)^2+4]} \bigg]
\end{align}
The third evaluates
\begin{equation}
\overleftarrow{{\rm Bub}}_3^{gg^*} = 4 (\ppp_1^2+1)(\ppp_2^2+1) \left[ \braket{\phi(p_1+p_2)\phi(-p_1-p_2)}^{(1)} + \braket{\phi(p_1-p_2)\phi(-p_1+p_2)}^{(1)} \right]
\end{equation}
This contribution is again UV finite and can be straightforwardly obtained from the computations of the last two sections.

In the backward kinematic configuration we can ascertain that the fourth and fifth bubble topologies and the tadpole ones vanish. This descends from the fact that these diagrams are all proportional to the tree-level amplitude and the latter vanishes for backward scattering, as verified in Section \ref{sec:tree}.

\subsection{Expression in terms of bubble integrals}

After tensor reduction the result in terms of bubbles is summarized in Tables \ref{tab:backpq} and \ref{tab:backp} in Appendix \ref{app:details}.
Remarkably the coefficients of all bubble integrals vanish apart from those with an external momentum flowing in the loop, yielding
\begin{equation}\label{eq:back}
\overleftarrow{\cal A}_{gg^*}^{(1)}(\theta_1,\theta_2)\Big|_{\pi} = \frac{\pi}{2}\, \frac{\cosh 2 \theta _1 \cosh 2 \theta _2 \left(\cosh 4 \theta _1 + \cosh 4 \theta _2 + 2\right)}{\cosh^2\left(\theta _1-\theta _2\right)} 
\end{equation}
Also, divergences and contributions proportional to $\log 2$ cancel out. Terms of lower transcendentality do so as well, provided the RSTW reduction is employed.

\section{External legs corrections}\label{sec:external}

External legs receive quantum corrections which must be taken into account.
This is carried out via the LSZ formula.
It entails considering the quantum corrections to the two-point function of the external particles, the gauge excitations in this case.
After re-summing the 1PI contributions to the all-loop propagator via a geometric series, one has to consider the residue $Z$ at the physical, quantum corrected pole.
This procedure is scheme dependent, as divergent bubble integrals with powers of loop momentum in the numerator show up.
In this section we review and revisit the computation of the two-point function of gauge excitations \cite{Giombi:2010bj} using the different schemes proposed in Section \ref{sec:tensor}.
The diagrams contributing to the one-loop correction of the two-point function evaluate
\begin{align}
\langle x(p)x^*(-p)\rangle^{(1)} &=
32\, \frac{p_1^2+1}{(p^2+2)^2}
\left[\int \frac{d^2l}{(2\pi)^2} \frac{l_1^2+1}{(l^2+2)((l+p)^2+4)}
-\frac{1}{2}\int \frac{d^2l}{(2\pi)^2}\frac{1}{l^2+4}\right. + \nonumber\\&
+ \left.\int \frac{d^2l}{(2\pi)^2} \frac{(l_0+p_0)l_0}{(l^2+1)((l+p)^2+1)}
-\int \frac{d^2l}{(2\pi)^2}\frac{l_1^2+1}{l^2+1}\right] 
\end{align}
Tensor reduction can be performed with different methods differing for scheme dependent terms.
We focus on the scheme independent part first.
This reads
\begin{align}\label{eq:x2pt}
\langle x(p)x^*(-p)\rangle^{(1)} &= \frac{\ppp^2+1}{(p^2+2)^2} \bigg[-\frac{8\left(p^4+8 p^2+4\right) \left(p^2-2 \ppp^2\right)}{p^4} \text{I}[2,4;p] + \nonumber\\& - \frac{8\left(p^4+4 p^2+4 \ppp^2+8\right)}{p^2} \text{I}[1,1;p] + \frac{8\left(p^2+2\right) \left(p^2-2 \ppp^2\right)}{p^4} \log 2 \bigg] = \nonumber\\& = 
\frac{2}{(p^2+2)^2}\, F^{(1)}(\ppp)
\end{align}
and coincides with the reduction performed with the RSTW scheme, which effectively removes all terms of lower transcendentality.
For completeness we also give the result in dimensional regularization, which features an extra term of lower transcendentality $\frac{20}{\pi} + \frac{16}{\pi\, p^2}$ inside the brackets.
The result \eqref{eq:x2pt} is re-summed as a geometric series giving the corrected two-point function
\begin{equation}
\langle x(p)x^*(-p)\rangle = \frac{1}{2g}\, \frac{2}{p^2+2 - \frac{1}{2g}\, F^{(1)}(\ppp)} + {\cal O}(g^{-2})
\end{equation}
The one-loop corrected dispersion relation can then be read off imposing that the denominator vanishes. 
Perturbatively, this can be achieved expanding $F^{(1)}(\ppp)$ close to the mass shell where we have
\begin{align}
F^{(1)}(\ppp) &= \frac{1}{2} \left(\ppp^2+1\right)^2 + \frac{1}{8\pi} \left[ -4 \left(p^2+2\right) \left(\ppp^2+1\right)^2 + \pi  \left(p^2+2\right) \left(\ppp^2-5\right) \left(\ppp^2+1\right) + \right.\nonumber\\&\left. -4 \left(p^2+2\right) \left(\ppp^2+1\right)^2 \log 2 \right] + {\cal O}\left((p^2+2)^2\right)
\end{align}
In dimensional regularization an extra piece $\frac{6}{\pi}$ appears, affecting the order 0 term.
To perform such an expansion we have used
\begin{align}
& {\rm I}[2,4;-2] = \frac{1}{4\pi}\left( \frac{\pi}{8} + \frac{\pi-4}{32}(p^2+2) \right) + {\cal O}\left((p^2+2)^2\right) \nonumber\\&
{\rm I}[1,1;-2] = \frac{1}{4\pi}\left( \frac{\pi}{2} - \frac{1}{2}(p^2+2) \right) + {\cal O}\left((p^2+2)^2\right)
\end{align}
Using the result above, the one-loop dispersion relation reads
\begin{equation}
e(\ppp) = \sqrt{-\ppp^2-2} + \frac{(\ppp^2+1)^2}{8g\sqrt{-\ppp^2-2}} + {\cal O}(g^{-2})
\end{equation}
The residue at this pole can be extracted and gives
\begin{equation}\label{eq:wavefunction}
Z(\ppp) = 1 + \frac{\ppp^2+1}{16 \pi  g} \left[ \pi (\ppp^2 -5 ) - 4 (1 + \log 2) (\ppp^2 +1 )\right] + {\cal O}(g^{-2})
\end{equation}
This expression holds for all the regularization schemes we have considered in the paper. A term of lower transcendentality explicitly appears in it.

The wave function renormalization $Z$ is pivotal for the corrections to the external legs to be included via the LSZ formalism.
According to it we add to the amplitude computed by summing the Feynman diagrams of the last sections the tree-level amplitude multiplied by a factor $\sqrt{Z}$ for each external leg.
For the same helicity amplitude, from \eqref{eq:treeamplitude2}, this gives
\begin{align}\label{eq:sameLSZ}
& S_{gg}^{(1)}(\theta_1,\theta_2)\Big|_{\rm LSZ} = 
- \frac{\cosh ^2 \frac{\theta _1-\theta _2}{2} \left(\cosh 4 \theta _1 + \cosh 4 \theta _2 + 2\right)}{8 \pi  \left(\tanh 2 \theta _1 - \tanh 2 \theta _2\right)} \log 2 + \nonumber\\&
- \frac{\cosh ^2 \frac{\theta _1-\theta _2}{2} \left(\cosh 4 \theta _1 + \cosh 4 \theta _2 + 2\right)}{8 \pi  \left(\tanh 2 \theta _1 - \tanh 2 \theta _2\right)} + 
\nonumber\\&
+ \frac{\cosh 2 \theta _1 \cosh 2 \theta _2 \cosh^2 \frac{\theta _1-\theta _2}{2}}{16\sinh 2\left( \theta _1 - \theta _2\right)} \left[ \cosh ^2 2 \theta _1 + \cosh ^2 2 \theta _2 - 6 \left(\cosh 2 \theta _1 + \cosh 2 \theta _2\right)\right]
\end{align}
For the opposite helicity and forward kinematic, from \eqref{eq:treexxb}, we obtain
\begin{align}\label{eq:oppLSZ}
& S_{gg^*}^{(1)}(\theta_1,\theta_2)\Big|_{\rm LSZ} = 
- \frac{\sinh ^2 \frac{\theta _1-\theta _2}{2} \left(\cosh 4 \theta _1 + \cosh 4 \theta _2 + 2\right)}{8 \pi  \left(\tanh 2 \theta _1 - \tanh 2 \theta _2\right)} \log 2 + \nonumber\\&
-  \frac{\sinh ^2 \frac{\theta _1-\theta _2}{2} \left(\cosh 4 \theta _1 + \cosh 4 \theta _2 + 2\right)}{8 \pi  \left(\tanh 2 \theta _1 - \tanh 2 \theta _2\right)} + \nonumber\\&
+ \frac{\cosh 2 \theta _1 \cosh 2 \theta _2 \sinh^2 \frac{\theta _1-\theta _2}{2}}{16\sinh 2\left( \theta _1 - \theta _2\right)} \left[ \cosh ^2 2 \theta _1 + \cosh ^2 2 \theta _2 - 6 \left(\cosh 2 \theta _1 + \cosh 2 \theta _2\right)\right]
\end{align}
whereas in the backward case this contribution vanishes as the tree-level amplitude does so.

\section{Corrections to the gluon dispersion relation}\label{sec:treedisp}

When expressing the scattering factors as a function of the rapidities one has to take into account that the tree-level results were originally expressed in terms of energies and spatial momenta of the scattering particles and that the energy, as a function of momentum receives quantum corrections as well.
Therefore we have to add to the result from the previous sections the contribution from plugging the quantum dispersion relations in the tree-level diagrams.
This substitution has to be performed on the Jacobian factor \eqref{eq:Jacobian} as well in order to capture all the terms contributing to order $g^{-2}$.

For scattering of gluons of the same helicity this correction factor reads
\begin{align}
& S_{gg}^{(1)}(\theta_1,\theta_2)\Big|_{\rm disp} = - \frac{\cosh 2 \theta _1 \cosh 2 \theta _2}{32\, \sinh 2 \left(\theta _1- \theta _2\right)} \bigg[
4 \cosh \left(\theta _1-\theta _2\right) \cosh ^2\left(\theta _1+\theta _2\right) + \nonumber\\&
-6 \left(\cosh \left(\theta _1-\theta _2\right)+1\right) \left(\cosh 2 \theta _1 + \cosh 2 \theta _2\right) + 4 \left(\cosh \left(\theta _1-\theta _2\right)+1\right) \left(1-\cosh 2 \left(\theta _1+\theta _2\right) \right) + \nonumber\\&
+\frac{1}{2} \left(\cosh ^2 2 \theta _1 + \cosh ^2 2 \theta _2\right) \left(\frac{\cosh \theta _2}{\cosh \theta _1} + \frac{\cosh \theta _1}{\cosh \theta _2}\right) + \nonumber\\&
-\frac{\left(\cosh 2 \theta _1 + \cosh 2 \theta _2\right) \left(\cosh 2 \theta _1 - \cosh 2 \theta _2\right)^2}{4 \cosh \theta _1 \cosh \theta _2} 
+ \frac{\cosh 2 \theta _1 \cosh 2 \theta _2 \left(1-\cosh \left(\theta _1+\theta _2\right)\right)}{\cosh \theta _1 \cosh \theta _2 \cosh\left(\theta _1-\theta _2\right)}\bigg]
\end{align}
For opposite helicity and forward kinematics we find
\begin{align}
& S_{gg^*}^{(1)}(\theta_1,\theta_2)\Big|_{\rm disp} =-  \frac{\cosh 2 \theta _1 \cosh 2 \theta _2}{32\sinh 2\left( \theta _1- \theta _2\right)}  \bigg[4 \cosh \left(\theta _1-\theta _2\right) \cosh ^2\left(\theta _1+\theta _2\right) + \nonumber\\&
+\cosh \left(\theta _1-\theta _2\right) \left(\cosh ^2 2 \theta _1 + \cosh ^2 2 \theta _2 \right)
+6 \left(1-\cosh \left(\theta _1-\theta _2\right)\right) \left(\cosh 2 \theta _1 + \cosh 2 \theta _2\right) + \nonumber\\&
-4 \left(1-\cosh \left(\theta _1-\theta _2\right)\right) \left(1-\cosh 2 \left(\theta _1+\theta _2\right)\right) - \frac{\cosh 2 \theta _1 \cosh 2 \theta_2 \left(\cosh \left(\theta _1+\theta _2\right)+1\right)}{\cosh \theta _1 \cosh \theta_2 \cosh \left(\theta _1-\theta _2\right)} + \nonumber\\& - \frac{\sinh \left(\theta _1-\theta _2\right)}{\cosh \theta _1 \cosh \theta _2} \left(\sinh \theta _1 \cosh ^2 2 \theta _1 \cosh \theta _2 - \sinh \theta _2 \cosh \theta _1 \cosh ^2 2 \theta _2\right)\bigg]
\end{align}
Finally, for backward kinematics we obtain
\begin{equation}\label{eq:dispback}
\overleftarrow{S}_{gg^*}^{(1)}(\theta_1,\theta_2)\Big|_{\rm disp} =- \frac{\cosh 2 \theta _1 \cosh 2 \theta _2 \left(\cosh 4 \theta _1 + \cosh 4 \theta _2 + 2\right)}{64\, \sinh\left(\theta _1-\theta _2\right) \cosh^2\left(\theta _1-\theta _2\right)}
\end{equation}
In this last expression there is no contribution from the Jacobian, as it multiplies a vanishing tree-level factor. However, while the diagrams for this amplitude cancel each other after imposing the relativistic on-shell conditions on the energies, they no longer do so when the corrected dispersion relations are used. This explains the emergence of the factor \eqref{eq:dispback}.

\section{Final results}\label{sec:final}

We finally combine all partial results derived in the previous sections to obtain the full gluon-gluon scattering factors at next-to-leading order. We express them as a function of the hyperbolic rapidities $\theta_i$.
We first focus on the scheme independent part of maximal transcendentality and then analyse the scheme dependent piece.
\paragraph{Same helicity}
For gauge excitations of the same helicity we sum all contributions 
\begin{align}
S_{gg}^{(1)}(\theta_1,\theta_2) &= \frac{1}{32 \pi\, \sinh \left(\theta_1-\theta_2\right)} \left( {\cal A}_{gg}^{(1)}\Big|_{\rm log} + {\cal A}_{gg}^{(1)}\Big|_{\pi} + {\cal A}_{gg}^{(1)}\Big|_{\log{2}} + {\cal A}_{gg}^{(1)}\Big|_{\rm algebraic} \right) + \nonumber\\& + S_{gg}^{(1)}(\theta_1,\theta_2)\Big|_{\rm LSZ} + S_{gg}^{(1)}(\theta_1,\theta_2)\Big|_{\rm disp}
\end{align}
and the final one-loop amplitude reads (for the scheme independent part)
\begin{align}
S_{gg}^{(1)}(\theta_1,\theta_2) &=i\, \frac{\left(\cosh \left(\theta _1-\theta _2\right)+1\right)^2}{8 \left(\tanh 2 \theta _1-\tanh 2 \theta _2\right)^2} 
+ \frac{\left(\cosh \left(\theta _1-\theta _2\right)+1\right)}{8 \pi  \left(\tanh 2 \theta _1-\tanh 2 \theta _2\right)}\, 3 \log 2 + \nonumber\\&
-  \frac{\cosh 2 \theta _1 \cosh 2 \theta _2}{32\, \sinh 2\left(\theta _1-\theta _2\right)} 
\Bigg[
\cosh \left(\theta _1-\theta _2\right) \left(\cosh ^2 2 \theta _1+\cosh ^2 2 \theta _2\right) + \nonumber\\&
+\frac{\cosh 2 \theta _1 \cosh 2 \theta _2 \left(1-\cosh \left(\theta _1+\theta _2\right)\right)}{\cosh \theta _1\cosh \theta _2 \cosh\left(\theta _1-\theta _2\right)} +  \nonumber\\& 
-\frac{\sinh \left(\theta _1-\theta _2\right)}{\cosh \theta _1 \cosh \theta _2} \left(\sinh\theta _1 \cosh ^2 2 \theta _1 \cosh \theta _2 - \sinh \theta _2 \cosh \theta _1 \cosh ^2 2 \theta _2\right)\Bigg]
\end{align}
The result is in complete agreement with the integrability prediction \eqref{eq:predictionsame}.
As concerns scheme dependent terms, we remark that in the RSTW reduction these terms cancel out.
Namely, the contribution ${\cal A}_{gg}^{(1)}\Big|_{\rm algebraic}$ \eqref{eq:samealgebraic} cancels exactly against the lower transcendentality piece of the LSZ correction $S_{gg}^{(1)}\Big|_{\rm LSZ}$ \eqref{eq:sameLSZ}.
In other words the scheme of \cite{Roiban:2014cia} is capable of reproducing the integrability result exactly.

\paragraph{Opposite helicity forward}
For gauge excitations of the opposite helicity and forward kinematics we combine all contributions
\begin{align}
S_{gg^*}^{(1)}(\theta_1,\theta_2) &= \frac{1}{32 \pi\, \sinh \left(\theta_1-\theta_2\right)} \left( {\cal A}_{gg^*}^{(1)}\Big|_{\rm log} + {\cal A}_{gg^*}^{(1)}\Big|_{\pi} + {\cal A}_{gg^*}^{(1)}\Big|_{\log{2}} + {\cal A}_{gg^*}^{(1)}\Big|_{\rm algebraic} \right) + \nonumber\\& + S_{gg^*}^{(1)}(\theta_1,\theta_2)\Big|_{\rm LSZ} + S_{gg^*}^{(1)}(\theta_1,\theta_2)\Big|_{\rm disp}
\end{align}
and obtain for the scheme independent part
\begin{align}
S_{gg^*}^{(1)}(\theta_1,\theta_2) &=i\, \frac{\left(\cosh \left(\theta _1-\theta _2\right)-1\right)^2}{8 \left(\tanh 2 \theta _1 - \tanh 2 \theta _2\right)^2} + 
\frac{\left(\cosh \left(\theta _1-\theta _2\right)-1\right)}{8 \pi  \left(\tanh 2 \theta _1 - \tanh 2 \theta _2\right)}\, 3 \log 2 + \nonumber\\&
-  \frac{\cosh 2 \theta _1 \cosh 2 \theta _2}{32\, \sinh 2\left(\theta _1-\theta _2\right)} 
\Bigg[
\cosh \left(\theta _1-\theta _2\right) \left(\cosh ^2 2 \theta _1+\cosh ^2 2 \theta _2\right) + \nonumber\\&
-\frac{\cosh 2 \theta _1 \cosh 2 \theta _2 \left(1+\cosh \left(\theta _1+\theta _2\right)\right)}{\cosh \theta _1\cosh \theta _2 \cosh\left(\theta _1-\theta _2\right)} +  \nonumber\\& 
-\frac{\sinh \left(\theta _1-\theta _2\right)}{\cosh \theta _1 \cosh \theta _2} \left(\sinh\theta _1 \cosh ^2 2 \theta _1 \cosh \theta _2 - \sinh \theta _2 \cosh \theta _1 \cosh ^2 2 \theta _2\right)\Bigg]
\end{align}
which completely agrees with \eqref{eq:predictionopposite}.
As before, the RSTW scheme produces a precise cancellation of lower transcendentality terms agreeing with the integrability result.
In particular the contribution ${\cal A}_{gg^*}^{(1)}\Big|_{\rm algebraic}$ \eqref{eq:oppalgebraic} cancels the algebraic part of the LSZ correction $S_{gg^*}^{(1)}\Big|_{\rm LSZ}$ \eqref{eq:oppLSZ}.

\paragraph{Opposite helicity backward}
For gauge excitations of the opposite helicity and backward kinematics we observe that the  term proportional to $\pi$ surviving the Feynman diagrams \eqref{eq:back} is exactly cancelled against the opposite contribution \eqref{eq:dispback} coming from the corrections to the tree-level amplitude induced by the dispersion relation of the gluons
\begin{equation}
\overleftarrow{S}_{gg^*}^{(1)} = \frac{1}{32 \pi\, \sinh \left(\theta_1-\theta_2\right)}\, \overleftarrow{\cal A}_{gg^*}^{(1)}\Big|_{\pi} + \overleftarrow{S}_{gg^*}^{(1)}(\theta_1,\theta_2)\Big|_{\rm disp} = 0
\end{equation}
Therefore we conclude that this amplitude vanishes, in agreement with the integrability prediction.

\section{Conclusions}

In this paper we have computed the scattering factors for the gauge excitations on top of the GKP vacuum at one-loop order in the strong coupling expansion.
The latter theory is conjectured to be integrable, which allows to compute S-matrix elements exactly from the ABA equations.
This can be expanded at strong coupling and the leading \cite{Fioravanti:2015dma} and next-to-leading \cite{Belitsky:2015qla} order terms have been recently worked out within this approach.
We have reproduced the gluonic next-to-leading order results from perturbation theory in the worldsheet Lagrangian of the GKP fluctuations. 
We have used a standard computation in terms of Feynman diagrams.
After performing the integral reduction, the results can be expressed in terms of bubble integrals which can be straightforwardly evaluated.
We have compared the final expressions with the prediction based on integrability and have found perfect agreement.
This holds true for the scheme independent part of the amplitude.
For the scheme dependent terms we have found that a recently proposed framework for reduction of tensor integrals \cite{Roiban:2014cia} exactly reproduces the integrability result.

The scattering factors of GKP string excitations are crucial ingredients in the OPE program for scattering amplitudes of ${\cal N}=4$ SYM, since they constitute the fundamental building blocks of pentagon transitions.
In particular, the next-to-leading order terms are pivotal for pushing the computation of scattering amplitudes to one loop at strong coupling.
In this paper we have given the results provided by integrability for the gluon S-matrix the solid backup of a perturbative field-theoretical computation.

Scattering factors involving other GKP excitations are also important in the OPE program. It would be interesting to investigate whether these could also be studied perturbatively.
From previous experience and literature, it should be feasible (though probably requiring some more effort than the present paper) to determine the one-loop corrections to the gluon-fermion, gluon-meson, meson-meson and meson-fermion amplitudes.

The computational power for loop scattering amplitudes in field theory has been boosted by the advent of unitarity based techniques. This framework has been developed extensively in the realm of four-dimensional models, but has also been recently extended to two-dimensional models \cite{Engelund:2013fja,Bianchi:2013nra,Bianchi:2014rfa}.
It would be interesting to re-derive and extend the results of this paper via unitarity.

\section*{Acknowledgements}

We thank Benjamin Basso, Davide Fioravanti, Valentina Forini, Ben Hoare, Simone Piscaglia and Marco Rossi for very useful discussions. 
The work of LB is supported by Deutsche Forschungsgemeinschaft in Sonderforschungsbereich 676 ``Particles, Strings, and the Early Universe''.
The work of MB was supported in part by the Science and Technology Facilities Council Consolidated Grant ST/L000415/1 \emph{String theory, gauge theory \& duality}.

\vfill
\newpage

\appendix

\section{ABA results}
\label{app:integrability}
In this appendix we collect the results from the ABA description derived in \cite{Belitsky:2015qla}, which are relevant for the expansion of the one-loop scattering factors of GKP gluons.
The computation is carried out in the strong perturbative regime, where Bethe rapidities have been rescaled as $u = 2g\,\bar u$. We drop the bar to avoid clutter, but stress that the following rapidities are understood as the rescaled ones. 
In order to compare the scattering factors with a direct computation from the worldsheet Lagrangian, we express the formulae above in terms of the spatial momentum of the incoming particles, or equivalently in terms of hyperbolic rapidities (with the identification $\ppp_i=\sqrt{2} \sinh\theta_i$)
This entails the quantum relation between Bethe rapidity and particle momentum for gluons, which was spelled out in \cite{Basso:2010in}
\begin{equation}
u(\ppp) = \frac{\ppp\, \sqrt{\ppp^2+2}}{\ppp^2+1} + \frac{1}{4 \pi\, g}\, \frac{\ppp\, \sqrt{\ppp^2+2}}{\ppp^2+1} \left(\frac{\pi}{2}\, \frac{\ppp^2+1}{\ppp^2+2} - 3 \log 2 \right) + {\cal O}(g^{-2})
\end{equation}
In terms of hyperbolic rapidities it reads
\begin{equation}\label{eq:rap}
u(\theta) = \tanh 2 \theta + \frac{1}{8\pi\, g} \left(\pi\, \tanh \theta - 6 \log 2\, \tanh 2 \theta\right) + {\cal O}(g^{-2})
\end{equation}
Following \cite{Belitsky:2015qla} the S-matrix reads
\begin{align}
\label{eq:Smatrix}
S_{gg} (u_1, u_2) & =
s (u_1, u_2) S_{gg^*} (u_1, u_2) = \nonumber\\&
= s (u_1, u_2) \exp \left( - 2 i f^{(1)}_{gg} (u_1, u_2) + 2 i f^{(2)}_{gg} (u_1, u_2) \right)
\end{align}
with
\begin{equation}
s (u_1, u_2) = \frac{u_1-u_2+i}{u_1-u_2-i}
\end{equation}
and
\begin{align}
\label{eq:f}
f_{gg}^{(\alpha)} (u_1, u_2) 
= \frac{1}{16 g}
\Bigg\{ A^{(\alpha)}_{gg} (u_1, u_2) 
+ \frac{1}{4 g}
\left[ B^{(\alpha)}_{gg} (u_1, u_2) 
+ \frac{3 \ln2}{2 \pi}  C^{(\alpha)}_{gg} (u_1, u_2) 
\right] + {\cal O} (g^{-2})
\Bigg\}
\end{align}
The distributions entering the expressions for the quantities $A$, $B$ and $C$, which were derived in \cite{Belitsky:2015qla}, require regularization, which is done by taking the principal value $P$.
The relevant functions for the order $g^{-1}$ calculation read
\begin{align}
A^{(1)}_{gg} (u_1, u_2) & =
\frac{2 P}{ u_1 - u_2} + 2 \pi i \delta (u_1 + u_2)
\nonumber\\&
- \frac{P}{u_1 - u_2}
\left[ \left( \frac{1 - u_1}{1 + u_1} \right)^{1/4} \left( \frac{1 + u_2}{1 - u_2} \right)^{1/4}
+ \left( \frac{1 + u_1}{1 - u_1} \right)^{1/4}
\left( \frac{1 - u_2}{1 + u_2} \right)^{1/4} \right]
\nonumber\\&
- \frac{P}{u_1 + u_2}
\left[ \left( \frac{1 - u_1}{1 + u_1} \right)^{1/4}
\left( \frac{1 - u_2}{1 + u_2} \right)^{1/4}
+ \left( \frac{1 + u_1}{1 - u_1} \right)^{1/4} \left( \frac{1 + u_2}{1 - u_2} \right)^{1/4}
\right]\\
A^{(3)}_{gg} (u_1, u_2) & =
- \frac{2 i P}{ u_1 + u_2} - 2 \pi \delta (u_1 + u_2)
\nonumber\\&
+ \frac{P}{u_1 - u_2}
\left[ \left( \frac{1 - u_1}{1 + u_1} \right)^{1/4}
\left( \frac{1 + u_2}{1 - u_2} \right)^{1/4}
- \left( \frac{1 + u_1}{1 - u_1} \right)^{1/4}
\left( \frac{1 - u_2}{1 + u_2} \right)^{1/4}\right]\nonumber\\&
-\frac{P}{u_1 + u_2}\left[
\left( \frac{1 - u_1}{1 + u_1} \right)^{1/4}
\left( \frac{1 - u_2}{1 + u_2} \right)^{1/4}
-\left( \frac{1 + u_1}{1 - u_1} \right)^{1/4}
\left( \frac{1 + u_2}{1 - u_2} \right)^{1/4}\right]\\
A^{(4)}_{gg} (u_1, u_2) & =
- \frac{2 i P}{ u_1 + u_2} - 2 \pi \delta (u_1 + u_2)\nonumber\\& 
- \frac{P}{u_1 - u_2}\left[
\left( \frac{1 - u_1}{1 + u_1} \right)^{1/4}
\left( \frac{1 + u_2}{1 - u_2} \right)^{1/4}
- \left( \frac{1 + u_1}{1 - u_1} \right)^{1/4}
\left( \frac{1 - u_2}{1 + u_2} \right)^{1/4}\right]
\nonumber\\&
-\frac{P}{u_1 + u_2} \left[
\left( \frac{1 - u_1}{1 + u_1} \right)^{1/4}
\left( \frac{1 - u_2}{1 + u_2} \right)^{1/4}
- \left( \frac{1 + u_1}{1 - u_1} \right)^{1/4}
\left( \frac{1 + u_2}{1 - u_2} \right)^{1/4}\right]
\end{align}
For the order $g^{-2}$ the contributing functions are
\begin{align}
& B^{(1)}_{gg} (u_1, u_2) = \frac{P}{[(u_1 - u_2)^2]_+}
\left[
\left( \frac{1 - u_1}{1 + u_1} \right)^{1/4}
\left( \frac{1 + u_2}{1 - u_2} \right)^{1/4}
- \left( \frac{1 + u_1}{1 - u_1} \right)^{1/4}
\left( \frac{1 - u_2}{1 + u_2} \right)^{1/4}\right]\nonumber\\
&\qquad\qquad\quad
+ \frac{P}{[(u_1 + u_2)^2]_+} \left[
\left( \frac{1 - u_1}{1 + u_1} \right)^{1/4}
\left( \frac{1 - u_2}{1 + u_2} \right)^{1/4}
- \left( \frac{1 + u_1}{1 - u_1} \right)^{1/4}
\left( \frac{1 + u_2}{1 - u_2} \right)^{1/4}
+ 2 i \right]
\nonumber\\&\qquad
+ \frac{2 - u_1^2 - u_2^2}{4 (1- u_1^2) (1- u_2^2)}
\frac{P}{u_1 - u_2} \left[
\left( \frac{1 - u_1}{1 + u_1} \right)^{1/4}
\left( \frac{1 + u_2}{1 - u_2} \right)^{1/4}
+ \left( \frac{1 + u_1}{1 - u_1} \right)^{1/4}
\left( \frac{1 - u_2}{1 + u_2} \right)^{1/4}
\right]\nonumber\\&\qquad
+\frac{2 - u_1^2 - u_2^2}{4 (1- u_1^2) (1- u_2^2)}
\frac{P}{u_1 + u_2}
\left[
\left( \frac{1 - u_1}{1 + u_1} \right)^{1/4}
\left( \frac{1 - u_2}{1 + u_2} \right)^{1/4}
+\left( \frac{1 + u_1}{1 - u_1} \right)^{1/4}
\left( \frac{1 + u_2}{1 - u_2} \right)^{1/4}
\right]\nonumber\\&\qquad\qquad\quad \
- 2 \pi \delta^\prime (u_1 + u_2)\\
&
B^{(3)}_{gg} (u_1, u_2) =
- \frac{P}{[(u_1 - u_2)^2]_+}
\left[
\left( \frac{1 - u_1}{1 + u_1} \right)^{1/4}
\left( \frac{1 + u_2}{1 - u_2} \right)^{1/4}
+ \left( \frac{1 + u_1}{1 - u_1} \right)^{1/4}
\left( \frac{1 - u_2}{1 + u_2} \right)^{1/4}
- 2 \right]\nonumber\\&\qquad\qquad\quad
+\frac{P}{[(u_1 + u_2)^2]_+}
\left[
\left( \frac{1 - u_1}{1 + u_1} \right)^{1/4}
\left( \frac{1 - u_2}{1 + u_2} \right)^{1/4}
+\left( \frac{1 + u_1}{1 - u_1} \right)^{1/4}
\left( \frac{1 + u_2}{1 - u_2} \right)^{1/4}
\right]\nonumber\\&\qquad
- \frac{2 - u_1^2 - u_2^2}{4 (1- u_1^2) (1- u_2^2)}
\frac{P}{u_1 - u_2}
\left[ \left( \frac{1 - u_1}{1 + u_1} \right)^{1/4}
\left( \frac{1 + u_2}{1 - u_2} \right)^{1/4}
- \left( \frac{1 + u_1}{1 - u_1} \right)^{1/4}
\left( \frac{1 - u_2}{1 + u_2} \right)^{1/4}\right]\nonumber\\&\qquad
+ \frac{2 - u_1^2 - u_2^2}{4 (1- u_1^2) (1- u_2^2)}
\frac{P}{u_1 + u_2}
\left[ \left( \frac{1 - u_1}{1 + u_1} \right)^{1/4}
\left( \frac{1 - u_2}{1 + u_2} \right)^{1/4}
- \left( \frac{1 + u_1}{1 - u_1} \right)^{1/4}
\left( \frac{1 + u_2}{1 - u_2} \right)^{1/4}
\right]\nonumber\\&\qquad\qquad\quad 
- 2 \pi i \, \delta^\prime (u_1 + u_2)\\
&
B^{(4)}_{gg} (u_1, u_2) =
\frac{P}{[(u_1 - u_2)^2]_+}
\left[ \left( \frac{1 - u_1}{1 + u_1} \right)^{1/4}
\left( \frac{1 + u_2}{1 - u_2} \right)^{1/4}
+ \left( \frac{1 + u_1}{1 - u_1} \right)^{1/4}
\left( \frac{1 - u_2}{1 + u_2} \right)^{1/4}
- 2 \right]\nonumber\\&\qquad\qquad\quad 
+ \frac{P}{[(u_1 + u_2)^2]_+}
\left[\left( \frac{1 - u_1}{1 + u_1} \right)^{1/4}
\left( \frac{1 - u_2}{1 + u_2} \right)^{1/4}
+ \left( \frac{1 + u_1}{1 - u_1} \right)^{1/4}
\left( \frac{1 + u_2}{1 - u_2} \right)^{1/4}\right]\nonumber\\&\qquad
+ \frac{2 - u_1^2 - u_2^2}{4 (1- u_1^2) (1- u_2^2)}
\frac{P}{u_1 - u_2}
\left[ \left( \frac{1 - u_1}{1 + u_1} \right)^{1/4}
\left( \frac{1 + u_2}{1 - u_2} \right)^{1/4}
- \left( \frac{1 + u_1}{1 - u_1} \right)^{1/4}
\left( \frac{1 - u_2}{1 + u_2} \right)^{1/4}\right]\nonumber\\&\qquad
+ \frac{2 - u_1^2 - u_2^2}{4 (1- u_1^2) (1- u_2^2)}
\frac{P}{u_1 + u_2}
\left[\left( \frac{1 - u_1}{1 + u_1} \right)^{1/4}
\left( \frac{1 - u_2}{1 + u_2} \right)^{1/4}
- \left( \frac{1 + u_1}{1 - u_1} \right)^{1/4}
\left( \frac{1 + u_2}{1 - u_2} \right)^{1/4}
\right] \nonumber\\ &\qquad\qquad\quad 
+ 2 \pi i \, \delta^\prime (u_1 + u_2)
\end{align}
and
\begin{align}
C^{(1)}_{gg} (u_1, u_2) & =
\frac{1 + u_1 u_2}{(1 - u_1^2)(1 - u_2^2)}
\left[
\left( \frac{1 - u_1}{1 + u_1} \right)^{1/4}
\left( \frac{1 + u_2}{1 - u_2} \right)^{1/4}
- \left( \frac{1 + u_1}{1 - u_1} \right)^{1/4}
\left( \frac{1 - u_2}{1 + u_2} \right)^{1/4} \right]
\nonumber\\&
+ \frac{1 - u_1 u_2}{(1 - u_1^2)(1 - u_2^2)}
\left[\left( \frac{1 - u_1}{1 + u_1} \right)^{1/4}
\left( \frac{1 - u_2}{1 + u_2} \right)^{1/4}
- \left( \frac{1 + u_1}{1 - u_1} \right)^{1/4}
\left( \frac{1 + u_2}{1 - u_2} \right)^{1/4}
\right]\\
C^{(3)}_{gg} (u_1, u_2) & =
- \frac{1 + u_1 u_2}{(1 - u_1^2)(1 - u_2^2)}
\left[\left( \frac{1 - u_1}{1 + u_1} \right)^{1/4}
\left( \frac{1 + u_2}{1 - u_2} \right)^{1/4}
+ \left( \frac{1 + u_1}{1 - u_1} \right)^{1/4}
\left( \frac{1 - u_2}{1 + u_2} \right)^{1/4}
\right]\nonumber\\&
+ \frac{1 - u_1 u_2}{(1 - u_1^2)(1 - u_2^2)}
\left[\left( \frac{1 - u_1}{1 + u_1} \right)^{1/4}
\left( \frac{1 - u_2}{1 + u_2} \right)^{1/4}
+ \left( \frac{1 + u_1}{1 - u_1} \right)^{1/4}
\left( \frac{1 + u_2}{1 - u_2} \right)^{1/4}
\right]\\
C^{(4)}_{gg} (u_1, u_2) & =
\frac{1 + u_1 u_2}{(1 - u_1^2)(1 - u_2^2)}
\left[\left( \frac{1 - u_1}{1 + u_1} \right)^{1/4}
\left( \frac{1 + u_2}{1 - u_2} \right)^{1/4}
+ \left( \frac{1 + u_1}{1 - u_1} \right)^{1/4}
\left( \frac{1 - u_2}{1 + u_2} \right)^{1/4}\right]\nonumber\\&
+ \frac{1 - u_1 u_2}{(1 - u_1^2)(1 - u_2^2)}
\left[\left( \frac{1 - u_1}{1 + u_1} \right)^{1/4}
\left( \frac{1 - u_2}{1 + u_2} \right)^{1/4}
+ \left( \frac{1 + u_1}{1 - u_1} \right)^{1/4}
\left( \frac{1 + u_2}{1 - u_2} \right)^{1/4}
\right]
\end{align}

We then arrive at the following predictions for scattering of gluons of same and opposite helicity at strong coupling
\begin{align}\label{eq:theBigBelitsky}
& S_{gg}(u_1,u_2) = 1 + \nonumber\\&~~~~ + \frac{i}{4g \left(u_1-u_2\right)} \left[\left(\frac{1+u_1}{1-u_1}\right)^{1/4} \left(\frac{1-u_2}{1+u_2}\right)^{1/4} + \left(\frac{1-u_1}{1+u_1}\right)^{1/4} \left(\frac{1+u_2}{1-u_2}\right)^{1/4} + 2\right] + \nonumber\\&~~~~ +
\frac{1}{g^2}\, \left\{- \frac{1}{32 \left(u_1-u_2\right)^2} \left[\left(\frac{1+u_1}{1-u_1}\right)^{1/4} \left(\frac{1-u_2}{1+u_2}\right)^{1/4} + \left(\frac{1-u_1}{1+u_1}\right)^{1/4} \left(\frac{1+u_2}{1-u_2}\right)^{1/4} + 2\right]^2 + \right.\nonumber\\&~~~~
+ \frac{i\left(\frac{1}{u_1^2-1} + \frac{1}{u_2^2-1}\right)}{64 \left(u_1-u_2\right)} \left[\left(\frac{1+u_1}{1-u_1}\right)^{1/4} \left(\frac{1-u_2}{1+u_2}\right)^{1/4} + \left(\frac{1-u_1}{1+u_1}\right)^{1/4} \left(\frac{1+u_2}{1-u_2}\right)^{1/4} \right] +  \nonumber\\&~~~~
+ \frac{i}{16} \left[\left(\frac{1+u_1}{1-u_1}\right)^{1/4} \left(\frac{1-u_2}{1+u_2}\right)^{1/4} - \left(\frac{1-u_1}{1+u_1}\right)^{1/4} \left(\frac{1+u_2}{1-u_2}\right)^{1/4} \right] \frac{3 \left(1 + u_1 u_2\right) \log 2}{2 \pi  \left(u_1^2-1\right) \left(u_2^2-1\right)} + 
\nonumber\\&~~~~ \left.
+ \frac{i}{16\left(u_1-u_2\right)^2} \left[\left(\frac{1+u_1}{1-u_1}\right)^{1/4} \left(\frac{1-u_2}{1+u_2}\right)^{1/4} - \left(\frac{1-u_1}{1+u_1}\right)^{1/4} \left(\frac{1+u_2}{1-u_2}\right)^{1/4} \right]
\right\} + {\cal O}(g^{-3}) \\
& S_{gg^\ast}(u_1,u_2) = 1 +  \nonumber\\&~~~~ + \frac{i}{4g \left(u_1-u_2\right)} \left[\left(\frac{1+u_1}{1-u_1}\right)^{1/4} \left(\frac{1-u_2}{1+u_2}\right)^{1/4} + \left(\frac{1-u_1}{1+u_1}\right)^{1/4} \left(\frac{1+u_2}{1-u_2}\right)^{1/4} - 2\right] +  \nonumber\\&~~~~ +
\frac{1}{g^2}\, \left\{ -\frac{1}{32 \left(u_1-u_2\right)^2} \left[\left(\frac{1+u_1}{1-u_1}\right)^{1/4} \left(\frac{1-u_2}{1+u_2}\right)^{1/4} + \left(\frac{1-u_1}{1+u_1}\right)^{1/4} \left(\frac{1+u_2}{1-u_2}\right)^{1/4} - 2\right]^2 + \right.\nonumber\\&~~~~
+ \frac{i\left(\frac{1}{u_1^2-1} + \frac{1}{u_2^2-1}\right)}{64 \left(u_1-u_2\right)} \left[\left(\frac{1+u_1}{1-u_1}\right)^{1/4} \left(\frac{1-u_2}{1+u_2}\right)^{1/4} + \left(\frac{1-u_1}{1+u_1}\right)^{1/4} \left(\frac{1+u_2}{1-u_2}\right)^{1/4} \right] + \nonumber\\&~~~~ 
+ \frac{i}{16} \left[\left(\frac{1+u_1}{1-u_1}\right)^{1/4} \left(\frac{1-u_2}{1+u_2}\right)^{1/4} - \left(\frac{1-u_1}{1+u_1}\right)^{1/4} \left(\frac{1+u_2}{1-u_2}\right)^{1/4} \right] \frac{3 \left(1 + u_1 u_2\right) \log 2}{2 \pi  \left(u_1^2-1\right) \left(u_2^2-1\right)}
 + \nonumber\\&~~~~ \left.
+ \frac{i}{16\left(u_1-u_2\right)^2} \left[\left(\frac{1+u_1}{1-u_1}\right)^{1/4} \left(\frac{1-u_2}{1+u_2}\right)^{1/4} - \left(\frac{1-u_1}{1+u_1}\right)^{1/4} \left(\frac{1+u_2}{1-u_2}\right)^{1/4} \right] 
\right\} + {\cal O}(g^{-3})\label{eq:theBigBelitsky2}
\end{align}
Plugging the expression \eqref{eq:rap} for the Bethe rapidities in \eqref{eq:theBigBelitsky} and \eqref{eq:theBigBelitsky2} we recover the results reported in section \ref{sec:result}.

\section{Details on the one-loop computation}\label{app:details}

In this appendix we collect several technical details on the one-loop computation that we did not include in the main text. In particular we provide a series of tables with the results of the integral and tensor reduction of the single Feynman diagrams. As usual when dealing with Feynman diagrams the simplicity of the final result is not transparent in the intermediate steps, which, in turn, look quite involved. For the opposite helicity scattering in forward kinematics we also provide the list of all the Feynman diagrams which is very similar to the same helicity case and therefore was not considered in the main text.
\subsection{Tensor reduction for same helicity scattering}

We start collecting the results of tensor reduction for integrals emerging in the same helicity scattering.
Table \ref{tab:samepq} summarizes tensor reduction for integrals with momentum invariant $u$.
\newpage
\begin{table}[!ht]
\centering
\resizebox{!}{0.85\textheight}{
\begin{sideways}
\begin{tabular}{l|ccc}
 & ${\rm I}[2,2;u]$ & ${\rm I}[4,4;u]$ & ${\rm I}[1,1;u]$ \\[1ex]
 \hline\noalign{\smallskip\smallskip}
${\rm Box}^{gg}_b$ & $\frac{\cosh 2\theta_1\cosh 2\theta_2 \left(\cosh 2\left(\theta _1-\theta _2\right) + 3\right)}{\cosh^2\left(\theta _1-\theta _2\right)}$ & $8 \left(\cosh \left(\theta _1+\theta _2\right)-2\right)^2$ & 0 \\[1ex]
${\rm Box}^{gg}_f$ & 0 & 0 & $-8 \frac{\cosh ^2\frac{\theta _1+\theta _2}{2}}{\cosh\left(\theta _1-\theta _2\right)} \left(2 \cosh \left(\theta _1-\theta _2\right)+ \cosh \left(\theta _1+\theta _2\right)-1\right)$ \\[1ex]
${\rm Tri}^{gg}_1$ & 0 & $32 \left(\cosh \left(\theta _1+\theta _2\right)-2\right)$ & 0 \\[1ex]
${\rm Tri}^{gg}_2$ & 0 & $-8 \left(-4 \cosh \left(\theta _1+\theta _2\right)+\cosh 2 \left(\theta _1+\theta _2\right)+1\right)$ & 0 \\[1ex]
${\rm Tri}^{gg}_3$ & $\frac{8 \sinh ^2\frac{\theta _1-\theta _2}{2} \cosh 2\theta_1 \cosh 2\theta_2}{\cosh^2\left(\theta _1-\theta _2\right)}$ &  & 0 \\[1ex]
${\rm Tri}^{gg}_4$ & 0 & 0 & $16 \frac{\cosh ^2\frac{\theta _1+\theta _2}{2}}{\cosh\left(\theta _1-\theta _2\right)} \left(2 \cosh \left(\theta _1-\theta _2\right)+\cosh \left(\theta _1+\theta _2\right)-1\right)$ \\[1ex]
${\rm Bub}^{gg}_1$ & 0 & 32 & 0 \\[1ex]
${\rm Bub}^{gg}_2$ & 0 & $-32 \cosh \left(\theta _1+\theta _2\right)$ & 0 \\[1ex]
${\rm Bub}^{gg}_3$ & $\frac{4 \cosh 2\theta_1 \cosh 2\theta_2}{\cosh^2\left(\theta _1-\theta _2\right)}$ & $8 \cosh ^2\left(\theta _1+\theta _2\right)$ & $-8 \frac{\cosh ^2\frac{\theta _1+\theta _2}{2}}{\cosh\left(\theta _1-\theta _2\right)} \left(2 \cosh \left(\theta _1-\theta _2\right)+\cosh \left(\theta _1+\theta _2\right)-1\right)$ \\[1ex]
\hline\noalign{\smallskip\smallskip}
total & $\frac{8 \cosh 2 \theta _1 \cosh 2 \theta _2 \cosh ^4 \frac{\theta _1-\theta _2}{2}}{\cosh^2\left(\theta _1-\theta _2\right)}$ & 0 & 0 
\end{tabular}
\end{sideways}}
\caption{Table of coefficients for bubbles with momentum $u$ for same helicity scattering.}
\label{tab:samepq}
\end{table}
The results for tensor reduction of bubbles with invariant $p^2=-2$ are given in Table \ref{tab:samep}.
\begin{table}[!ht]
\centering
\resizebox{!}{0.85\textheight}{
\begin{sideways}
\begin{tabular}{l|cc}
 & ${\rm I}[2,4;-2]$ & ${\rm I}[1,1;-2]$ \\[1ex]
 \hline\noalign{\smallskip}
${\rm Box}^{gg}_b$ & \parbox[c]{10cm}{
\begin{align*} 
& \Big[(2 \cosh 4 \theta _1+3 \cosh \left(\theta _1-5 \theta _2\right)+8 \cosh \left(2 \theta _1-2 \theta _2\right) + \\& +4 \cosh \left(\theta _1-\theta _2\right)+ 3 \cosh \left(5 \theta _1-\theta _2\right)+2 \cosh 4 \theta _2 + \\& + 7 \cosh \left(3 \theta _1+\theta _2\right)+7 \cosh \left(\theta _1+3 \theta _2\right)+12\Big] \text{sech}^2\left(\theta _1-\theta _2\right)
\end{align*}
} & 0 \\[1ex]
${\rm Box}^{gg}_f$ & 0 & $-4 \left(\cosh 2 \left(\theta _1+\theta _2\right)+\cosh \left(3 \theta _1+\theta _2\right)+\cosh \left(\theta _1+3 \theta _2\right)+3\right)$ \\[1ex]
${\rm Tri}^{gg}_2$ & \parbox[c]{11cm}{
\begin{align*}
& 2 \Big[3 \cosh \left(\theta _1-5 \theta _2\right)-4 \cosh \left(\theta _1-\theta _2\right) + \\& + 3 \cosh \left(5 \theta _1-\theta _2\right)
+4 \cosh 2 \left(\theta _1+\theta _2\right) + 7 \cosh \left(3 \theta _1+\theta _2\right) + \\& + 7 \cosh \left(\theta _1+3 \theta _2\right) - 4\Big] \text{sech}\left(\theta _1-\theta _2\right)
\end{align*}} & 0 \\[1ex]
${\rm Tri}^{gg}_3$ & \parbox[c]{11cm}{
\begin{align*}
& -6 \cosh 4 \theta _1-2 \Big[8 \cosh 2\theta_2+3 \cosh 4 \theta _2+2 \left(8 \cosh \left(\theta _1+\theta _2\right)+ \right.\\& +4 \cosh 2 \left(\theta _1+\theta _2\right) +
\text{sech}^2\left(\theta _1-\theta _2\right) \left(-2 \sinh ^2\left(\theta _1+\theta _2\right)\right.\\&\left.\left. +2 \cosh 2\theta_1+\cosh \left(4 \theta _1-2 \theta _2\right)+\cosh 2\theta_2\right)+5\right)\Big]
\end{align*}} & 0 \\[1ex]
${\rm Tri}^{gg}_4$ & 0 & \parbox[c]{10cm}{
\begin{align*}
& -2 \Big[\cosh 4 \theta _1-4 \cosh 2\theta_2+\cosh 4 \theta _2 + \\& - 8 \cosh \left(\theta _1+\theta _2\right) +4 \cosh \left(3 \theta _1+\theta _2\right)+ \\& -4 \cosh 2\theta_1 \left(\cosh 2\theta_1 \text{sech}\left(\theta _1-\theta _2\right)+1\right)-10\Big]
\end{align*}} \\[1ex]
${\rm Bub}^{gg}_4$ & $32 \left(\cosh 2\theta_1+\cosh 2\theta_2+2 \cosh \left(\theta _1+\theta _2\right)\right)$ & 0 \\[1ex]
${\rm Bub}^{gg}_5$ & 0 & $-12 \left(\cosh 2\theta_1+\cosh 2\theta_2+2 \cosh \left(\theta _1+\theta _2\right)\right)$ 
\end{tabular}
\end{sideways}}
\caption{Table of coefficients for bubbles with momentum $p^2=-2$ for same helicity scattering.}
\label{tab:samep}
\end{table}
\newpage
Table \ref{tab:same0} provides the coefficients of the scalar bubbles with vanishing inflowing momentum.
\begin{table}[!ht]
\centering
\resizebox{!}{0.85\textheight}{
\begin{sideways}
\begin{tabular}{l|ccc}
 & ${\rm I}[2,2,0]$ & ${\rm I}[4,4,0]$ & ${\rm I}[1,1,0]$ \\[1ex]
 \hline\noalign{\smallskip\smallskip}
${\rm Box}^{gg}_b$ & $-4 \left(\cosh \left(3 \theta _1+\theta _2\right)+\cosh \left(\theta _1+3 \theta _2\right)\right)$ & $-16 \left(\cosh 2\theta _1+\cosh 2\theta _2-2\right)$ & 0 \\[1ex]
${\rm Box}^{gg}_f$ & 0 & 0 & \parbox[c]{11cm}{
\begin{align*}
&
4 \Big[\sinh \theta _1 \sinh 3 \theta _2 - \cosh 2\theta _2 + \\& 
+\cosh \theta _1 \left(\cosh 3 \theta _2 -2 \cosh \theta _1\right)
+ \frac{\cosh ^2 2 \theta _1}{\cosh\left(\theta _1-\theta _2\right)}\Big]
\end{align*}} \\[1ex]
${\rm Tri}^{gg}_1$ & 0 & $16 \left(\cosh 2\theta _1+\cosh 2\theta _2-4\right)$ & 0 \\[1ex]
${\rm Tri}^{gg}_2$ & 0 & $-4 \left(\cosh 4\theta _1+\cosh 4\theta _2+2\right)$ & 0 \\[1ex]
${\rm Tri}^{gg}_3$ & 0 & 0 & 0 \\[1ex]
${\rm Tri}^{gg}_4$ & 0 & 0 & $2 \left(2 \cosh 2\theta _1+\cosh 4\theta _1+2 \cosh 2\theta _2+\cosh 4\theta _2+6\right)$ \\[1ex]
${\rm Bub}^{gg}_1$ & 0 & 32 & 0 \\[1ex]
${\rm Bub}^{gg}_2$ & 0 & 0 & 0 \\[1ex]
${\rm Bub}^{gg}_3$ & $2$ & $4$ & $-6$ \\[1ex]
\hline\noalign{\smallskip\smallskip}
total & $-2 \left(2 \cosh \left(3 \theta _1+\theta _2\right)+2 \cosh \left(\theta _1+3 \theta _2\right)-1\right)$ & $-4 \left(\cosh 4\theta _1+\cosh 4\theta _2+1\right)$ & \parbox[c]{11cm}{
\begin{align*}
& \Big[ 2 \cosh 4\theta _1+\cosh \left(\theta _1-5 \theta _2\right)+2 \cosh \left(\theta _1-\theta _2\right) + \\&
+\cosh \left(5 \theta _1-\theta _2\right)+2 \cosh 4\theta _2+2 \cosh 2 \left(\theta _1+\theta _2\right) + \\&
+\cosh \left(3 \theta _1+\theta _2\right)+\cosh \left(\theta _1+3 \theta _2\right)+2\Big] \text{sech}\left(\theta _1-\theta _2\right)
\end{align*}}
\end{tabular}
\end{sideways}}
\caption{Table of coefficients for bubbles with momentum 0 for same helicity scattering.}
\label{tab:same0}
\end{table}
\newpage

\subsection{One-loop opposite helicity forward scattering}

\subsubsection*{Feynman diagrams}
The computation of the opposite helicity scattering in forward kinematics can be straightforwardly derived from the same helicity case, using crossing relations.
Hence its description was cut short in Section \ref{sec:opphelicityf}.
For completeness, we report here the list of all relevant diagrams for scattering of two gluons with opposite helicity and forward kinematics.
We group them according to their topology.

\paragraph{Boxes}
The box diagrams are depicted in Figure \ref{fig:boxesF}.
\FIGURE[h]{
\centering
\includegraphics[width=1.\textwidth]{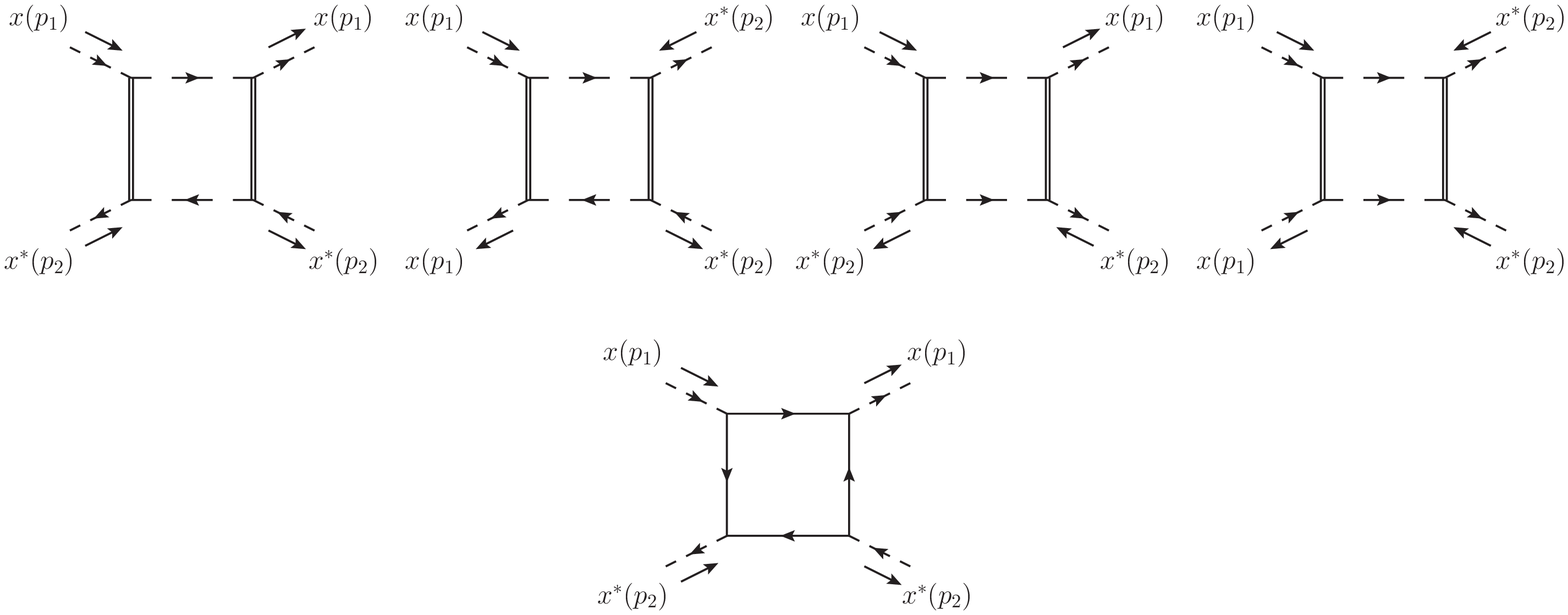}
\caption{Box diagrams in the opposite helicity process and for forward kinematics.}
\label{fig:boxesF}
}
Again four contractions are possible in the bosonic case which read
\begin{align}\label{eq:bosonboxF}
\overrightarrow{{\rm Box}}_b^{gg^*} &= 64(\ppp_1^2+1)(\ppp_2^2+1) \bigg\{
\int \frac{d^2l}{(2\pi)^2} \frac{[(l_1-\ppp_1)^2+1][(l_1+\ppp_2)^2+1]}{(l^2+4)^2[(l-p_1)^2+2][(l+p_2)^2+2]} + \nonumber\\&
+ \frac{[(l_1+\ppp_1)^2+1][(l_1+\ppp_2)^2+1]}{(l^2+4)^2[(l+p_1)^2+2][(l+p_2)^2+2]} + \nonumber\\&
+ \frac{(l_1^2+1)[(l_1-\ppp_1+\ppp_2)^2+1]}{(l^2+2)[(l-p_1)^2+4][(l-p_1+p_2)^2+2][(l+p_2)^2+4]} + \nonumber\\&
+ \frac{(l_1^2+1)^2}{(l^2+2)^2[(l-p_1)^2+4][(l+p_2)^2+4]} \bigg\}
\end{align}
The first two diagrams are the same as in the same helicity case. The last two are obtained from the analogous for same helicity scattering by sending, e.g., $p_1\to -p_1$.
The fermionic box algebra gives
\begin{equation}
\overrightarrow{{\rm Box}}_f^{gg^*} = -64 (\ppp_1^2+1)(\ppp_2^2+1)  \int\frac{d^2l}{(2\pi)^2} \frac{l_0^2 (l_0-e_1) (l_0+e_2)}{[l^2+1]^2[(l-p_1)^2+1][(l+p_2)^2+1]}
\end{equation}
and is again obtained from the same helicity fermion box by $p_1\to -p_1$.
As before, box diagrams are finite by power counting.

\paragraph{Triangles}
Triangle diagrams are shown in Figure \ref{fig:trianglesF} in a particular configuration.
\FIGURE{
\centering
\includegraphics[width=1.\textwidth]{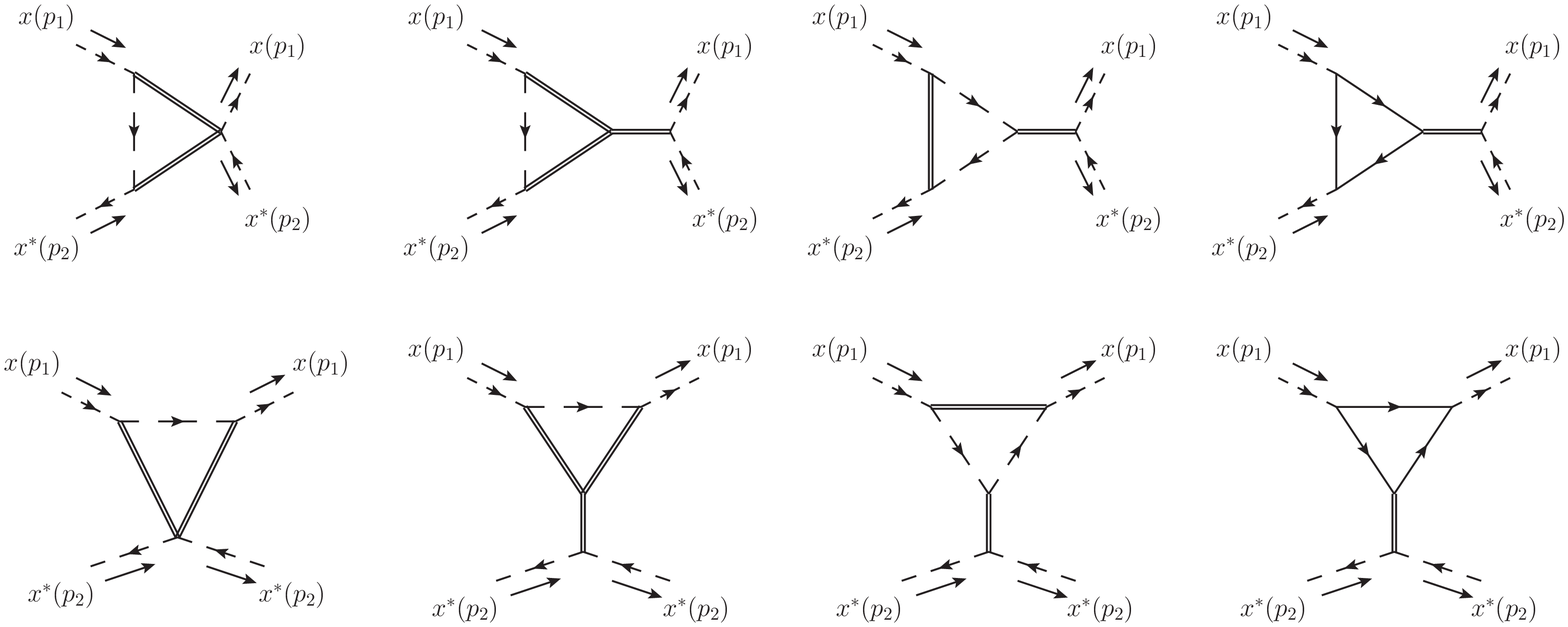}
\caption{Triangle diagrams in the opposite helicity process and for forward kinematics.}
\label{fig:trianglesF}
}
There is an additional permutation, reflecting the diagrams of the first line along a vertical axis and those of the second line along a horizontal one.
For the former this amounts to a factor of 2, for the latter to a $p_1\leftrightarrow p_2$ exchange. The results for the triangle diagrams follow.
The first bosonic triangle reads
\begin{align}
\overrightarrow{{\rm Tri}}_1^{gg^*} &= - 128(\ppp_1^2+1)(\ppp_2^2+1)  \int\frac{d^2l}{(2\pi)^2} \frac{(l_1^2+1)}{(l^2+2)[(l-p_1)^2+4][(l+p_2)^2+4]} + \nonumber\\&
- 64(\ppp_1^2+1)(\ppp_2^2+1)  \int\frac{d^2l}{(2\pi)^2} \frac{(l_1^2+1)}{(l^2+2)[(l+p_1)^2+4]^2} + \nonumber\\&
- 64(\ppp_1^2+1)(\ppp_2^2+1)  \int\frac{d^2l}{(2\pi)^2} \frac{(l_1^2+1)}{(l^2+2)[(l+p_2)^2+4]^2}
\end{align}
and is finite.
The second triangle evaluates
\begin{align}
\overrightarrow{{\rm Tri}}_2^{gg^*} &= - \frac{64(\ppp_1^2+1)(\ppp_2^2+1)}{(p_1+p_2)^2 + 4}  \int\frac{d^2l}{(2\pi)^2} \frac{(l_1^2+1)[l_0^2+l_0(e_2-e_1)+e_1^2+e_2^2+e_1e_2 - (t\leftrightarrow s)]}{(l^2+2)[(l-p_1)^2+4][(l+p_2)^2+4]} + \nonumber\\&
- \frac{32(\ppp_1^2+1)(\ppp_2^2+1)}{4}  \int\frac{d^2l}{(2\pi)^2} \frac{(l_1^2+1)[(l_0+e_1)^2-(l_1+\ppp_1)^2]}{(l^2+2)[(l+p_1)^2+4]^2} + \nonumber\\&
- \frac{32(\ppp_1^2+1)(\ppp_2^2+1)}{4}  \int\frac{d^2l}{(2\pi)^2} \frac{(l_1^2+1)[(l_0+e_2)^2-(l_1+\ppp_2)^2]}{(l^2+2)[(l+p_2)^2+4]^2}
\end{align}
and the integrals with four powers of loop momentum are UV divergent.
The last bosonic topology gives
\begin{align}
\overrightarrow{{\rm Tri}}_3^{gg^*} & = \frac{128(\ppp_1^2+1)(\ppp_2^2+1)}{(p_1+p_2)^2 + 4}  \int\frac{d^2l}{(2\pi)^2} \frac{[(l_1-\ppp_1)^2+1][(l_1+\ppp_2)^2+1]}{(l^2+4)[(l-p_1)^2+2][(l+p_2)^2+2]} + \nonumber\\&
+ \frac{64(\ppp_1^2+1)(\ppp_2^2+1)}{4}  \int\frac{d^2l}{(2\pi)^2} \frac{[(l_1+\ppp_1)^2+1]^2}{(l^2+4)[(l+p_1)^2+2]^2} + \nonumber\\& + \frac{64(\ppp_1^2+1)(\ppp_2^2+1)}{4}  \int\frac{d^2l}{(2\pi)^2} \frac{[(l_1+\ppp_2)^2+1]}{(l^2+4)[(l+p_2)^2+2]^2}
\end{align}
and is again divergent.
Finally there is a fermion loop diagram
\begin{align}
\overrightarrow{{\rm Tri}}_4^{gg^*} &= \frac{128(\ppp_1^2+1)(\ppp_2^2+1)}{(p_1+p_2)^2 + 4}  \int\frac{d^2l}{(2\pi)^2} \frac{l_0(l_0-e_1)[(l_1+\ppp_2)^2+1]+l_0(l_0+e_2)[(l_1-\ppp_1)^2+1]}{(l^2+1)[(l-p_1)^2+1][(l+p_2)^2+1]} + \nonumber\\&
+ \frac{128(\ppp_1^2+1)(\ppp_2^2+1)}{4}  \int\frac{d^2l}{(2\pi)^2} \frac{l_0(l_0+e_1)[(l_1+\ppp_1)^2+1]}{(l^2+1)[(l+p_1)^2+1]^2} + \nonumber\\& + \frac{128(\ppp_1^2+1)(\ppp_2^2+1)}{4}  \int\frac{d^2l}{(2\pi)^2} \frac{l_0(l_0+e_2)[(l_1+\ppp_2)^2+1]}{(l^2+1)[(l+p_2)^2+1]^2}
\end{align}
which is divergent.

\paragraph{Bubbles}
Bubble diagrams are shown in Figure \ref{fig:bubblesF}.
Again there are obvious permutations (affecting all diagrams but the first and the third) which have to be performed to include all combinations.
\FIGURE{
\centering
\includegraphics[width=1.\textwidth]{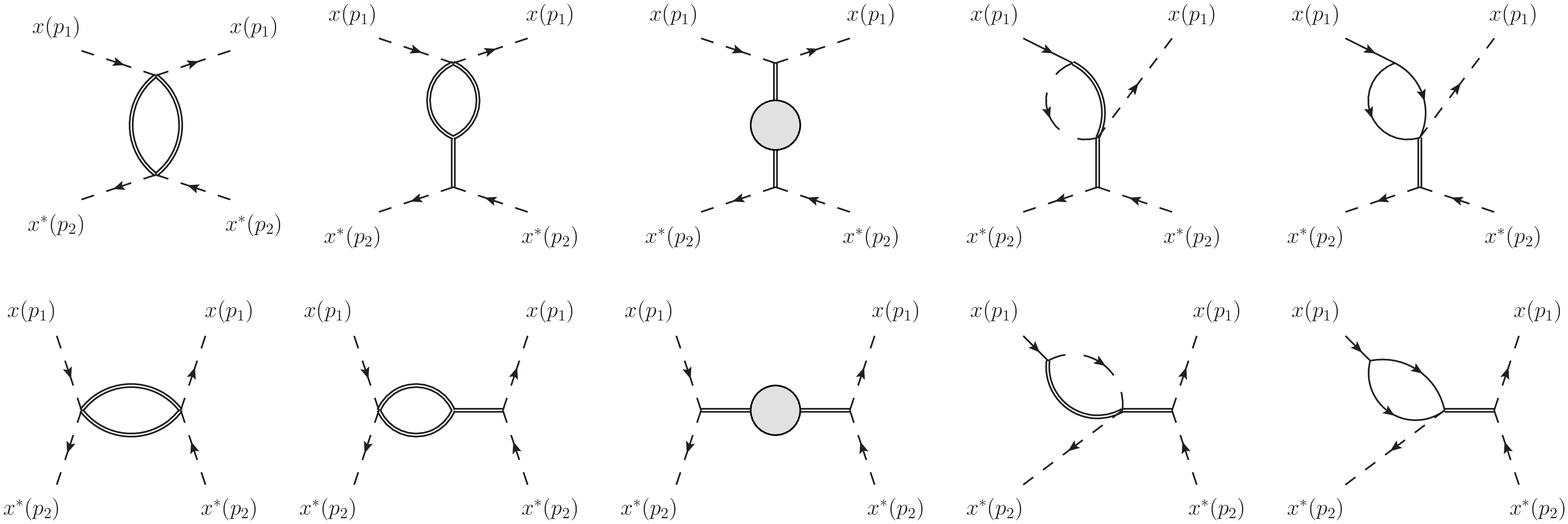}
\caption{Bubble diagrams in the opposite helicity process and for forward kinematics.}
\label{fig:bubblesF}
}
In particular, the second topology acquires an additional factor 2 (in each of the channels), whereas the last two topologies have to be multiplied by 4, stemming for the four possible external legs where the bubble is inserted.
These factors have already been included in the results which follow.
The first bubble diagram yields
\begin{equation}
\overrightarrow{{\rm Bub}}_1^{gg^*} = 32 (\ppp_1^2+1)(\ppp_2^2+1) \left[ {\rm I}[4,4;0]+{\rm I}[4,4;s] \right]
\end{equation}
and is finite.
The second bubble evaluates
\begin{align}
\overrightarrow{{\rm Bub}}_2^{gg^*} &= 32 (\ppp_1^2+1)(\ppp_2^2+1) \bigg[ \frac14 \int \frac{d^2l}{(2\pi)^2} \frac{l_0^2-l_1^2}{(l^2+4)^2} + \nonumber\\& 
+ \frac{1}{(p_1+p_2)^2+4} \int \frac{d^2l}{(2\pi)^2} \frac{l_0^2+l_0(e_1+e_2)+(e_1+e_2)^2 - (t \leftrightarrow s)}{(l^2+4)^2[(l+p_1+p_2)^2+4]} \bigg]
\end{align}
and as before it would be UV divergent by power counting, were it not for a cancellation of divergences thanks to $t$, $s$ antisymmetry.
This also forces the first integral to vanish, leaving
\begin{align}
\overrightarrow{{\rm Bub}}_2^{gg^*} &= \frac{32 (\ppp_1^2+1)(\ppp_2^2+1)}{(p_1+p_2)^2+4} \int \frac{d^2l}{(2\pi)^2} \frac{l_0^2 + l_0(e_2-e_1) + e_1 e_2 + e_1^2 + e_2^2 - (t \leftrightarrow s)}{[(l-p_1)^2+4][(l+p_2)^2+4]}
\end{align}
The third bubble is constructed from the one-loop correction to the meson propagator
\begin{equation}
\overrightarrow{{\rm Bub}}_3^{gg^*} = 4 (\ppp_1^2+1)(\ppp_2^2+1) \left[ \braket{\phi(0)\phi(0)}^{(1)} + \braket{\phi(p_1+p_2)\phi(-p_1-p_2)}^{(1)} \right]
\end{equation}
The same steps as above can be carried out to evaluate this contribution explicitly from the two-point function of \cite{Giombi:2010bj}.
The last two bubbles give
\begin{align}
\overrightarrow{{\rm Bub}}_4^{gg^*} & = -128 (\ppp_1^2+1)(\ppp_2^2+1) \left[ \frac{1}{(p_1+p_2)^2+4} + \frac14 \right] \bigg[ \int \frac{d^2l}{(2\pi)^2} \frac{l_1^2+1}{(l^2+2)[(l+p_1)^2+4]} + \nonumber\\&
+ \int \frac{d^2l}{(2\pi)^2} \frac{l_1^2+1}{(l^2+2)[(l+p_2)^2+4]} \bigg]
\end{align}
and
\begin{align}
\overrightarrow{{\rm Bub}}_5^{gg^*} &= -96 (\ppp_1^2+1)(\ppp_2^2+1) \left[ \frac{1}{(p_1+p_2)^2+4} + \frac14 \right] \bigg[ \int \frac{d^2l}{(2\pi)^2} \frac{l_0(l_0+e_1)}{(l^2+1)[(l+p_1)^2+1]} + \nonumber\\&
+ \int \frac{d^2l}{(2\pi)^2} \frac{l_0(l_0+e_2)}{(l^2+1)[(l+p_2)^2+1]} \bigg]
\end{align}
respectively.

\paragraph{Tadpoles}
The tadpole diagrams for the opposite helicity case are shown in Figure \ref{fig:tadpoleF} up to permutations.
\FIGURE{
\centering
\includegraphics[width=1.\textwidth]{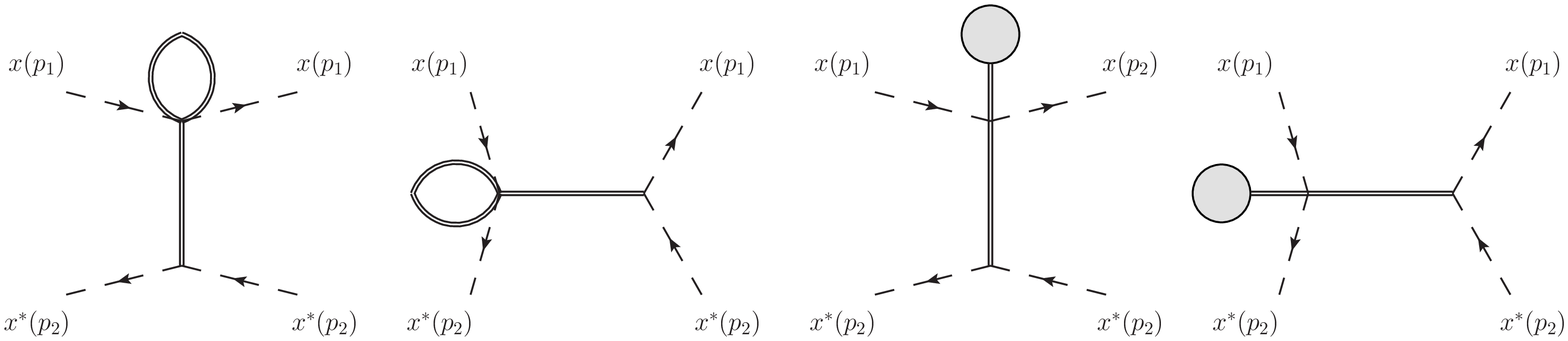}
\caption{Tadpole diagrams in the opposite helicity process and for forward kinematics.}
\label{fig:tadpoleF}
}
The first topology features a meson tadpole
\begin{equation}
\overrightarrow{{\rm Tad}}_1^{gg^*} = 64 (\ppp_1^2+1)(\ppp_2^2+1) \left[ \frac{1}{(p_1+p_2)^2+4} + \frac14 \right] {\rm I}[4]
\end{equation}
The second graph evaluates
\begin{equation}
\overrightarrow{{\rm Tad}}_2^{gg^*} = 64 (\ppp_1^2+1)(\ppp_2^2+1) \left[ \frac{1}{(p_1+p_2)^2+4} + \frac14 \right] {\rm I}[1]
\end{equation}
\newpage
\subsubsection*{Tensor reduction for opposite helicity forward scattering}

After tensor reduction the coefficients of bubbles with invariant $s$ are collected in Table \ref{tab:forpq}.
\begin{table}[!ht]
\centering
\resizebox{!}{0.85\textheight}{
\begin{sideways}
\begin{tabular}{l|ccc}
 & ${\rm I}[2,2;s]$ & ${\rm I}[4,4;s]$ & ${\rm I}[1,1;s]$ \\[1ex]
 \hline\noalign{\smallskip}
$\overrightarrow{{\rm Box}}^{gg^*}_b$ & $\frac{\cosh 2 \theta _1 \cosh 2 \theta _2 \left(\cosh 2\left( \theta _1-\theta _2\right) + 3\right)}{\cosh^2\left(\theta _1-\theta _2\right)}$ & 8 $\left(\cosh \left(\theta _1+\theta _2\right)+2\right){}^2$ & 0 \\[1ex]
$\overrightarrow{{\rm Box}}^{gg^*}_f$ & 0 & 0 & $8 \sinh ^2\frac{\theta _1+\theta _2}{2} \left(\frac{\cosh \left(\theta _1+\theta _2\right)+1}{ \cosh\left(\theta _1-\theta _2\right)}+2\right)$ \\[1ex]
$\overrightarrow{{\rm Tri}}^{gg^*}_1$ & 0 & $-32 \left(\cosh \left(\theta _1+\theta _2\right)+2\right)$ & 0 \\[1ex]
$\overrightarrow{{\rm Tri}}^{gg^*}_2$ & 0 & $-8 \left(4 \cosh \left(\theta _1+\theta _2\right)+\cosh 2 \left(\theta _1+\theta _2\right)+1\right)$ & 0 \\[1ex]
$\overrightarrow{{\rm Tri}}^{gg^*}_3$ & $-\frac{8 \cosh 2 \theta _1 \cosh 2 \theta _2 \cosh ^2\frac{\theta _1-\theta _2}{2}}{ \cosh^2\left(\theta _1-\theta _2\right)}$ & 0 & 0 \\[1ex]
$\overrightarrow{{\rm Tri}}^{gg^*}_4$ & 0 & 0 & $-16 \sinh ^2\frac{\theta _1+\theta _2}{2} \left(\frac{\cosh \left(\theta _1+\theta _2\right)+1}{ \cosh\left(\theta _1-\theta _2\right)}+2\right)$ \\[1ex]
$\overrightarrow{{\rm Bub}}^{gg^*}_1$ & 0 & 32 & 0 \\[1ex]
$\overrightarrow{{\rm Bub}}^{gg^*}_2$ & 0 & $32 \cosh \left(\theta _1+\theta _2\right)$ & 0 \\[1ex]
$\overrightarrow{{\rm Bub}}^{gg^*}_3$ & $4\frac{\cosh 2 \theta _1 \cosh 2 \theta _2}{\cosh^2\left(\theta _1-\theta _2\right)}$ & $8 \cosh ^2\left(\theta _1+\theta _2\right)$ & $8 \sinh ^2\frac{\theta _1+\theta _2}{2} \left(\frac{\cosh \left(\theta _1+\theta _2\right)+1}{ \cosh\left(\theta _1-\theta _2\right)}+2\right)$ \\[1ex]
\hline\noalign{\smallskip}
total & $8\, \sinh ^4 \frac{\theta _1-\theta _2}{2}\, \frac{\cosh 2 \theta _1 \cosh 2 \theta _2}{\cosh^2\left(\theta _1-\theta _2\right)}$ & 0 & 0\\[1ex]
\end{tabular}
\end{sideways}}
\caption{Table of the coefficients of bubbles with momentum $s$ for opposite helicity scattering and forward kinematics.}
\label{tab:forpq}
\end{table}
\newpage
In Table \ref{tab:forp} we provide details of the tensor reduction of bubbles with momentum invariant $p^2=-2$.
\begin{table}[!ht]
\centering
\resizebox{!}{0.85\textheight}{
\begin{sideways}
\begin{tabular}{l|cc}
 & ${\rm I}[2,4;-2]$ & ${\rm I}[1,1;-2]$ \\[1ex]
 \hline\noalign{\smallskip}
$\overrightarrow{{\rm Box}}^{gg^*}_b$ & \parbox[c]{10cm}{
\begin{align*} 
& \Big[(2 \cosh 4 \theta _1-3 \cosh \left(\theta _1-5 \theta _2\right)+8 \cosh \left(2 \theta _1-2 \theta _2\right) + \\& -4 \cosh \left(\theta _1-\theta _2\right)- 3 \cosh \left(5 \theta _1-\theta _2\right)+2 \cosh 4 \theta _2 + \\& - 7 \cosh \left(3 \theta _1+\theta _2\right)-7 \cosh \left(\theta _1+3 \theta _2\right)+12\Big] \text{sech}^2\left(\theta _1-\theta _2\right)
\end{align*}
} & 0 \\[1ex]
$\overrightarrow{{\rm Box}}^{gg^*}_f$ & 0 & $4 \left(-\cosh 2 \left(\theta _1+\theta _2\right)+\cosh \left(3 \theta _1+\theta _2\right)+\cosh \left(\theta _1+3 \theta _2\right)-3\right)$ \\[1ex]
$\overrightarrow{{\rm Tri}}^{gg^*}_2$ & \parbox[c]{11cm}{
\begin{align*}
& 2 \Big[3 \cosh \left(\theta _1-5 \theta _2\right)-4 \cosh \left(\theta _1-\theta _2\right) + \\& + 3 \cosh \left(5 \theta _1-\theta _2\right)
-4 \cosh 2 \left(\theta _1+\theta _2\right) + 7 \cosh \left(3 \theta _1+\theta _2\right) + \\& + 7 \cosh \left(\theta _1+3 \theta _2\right) + 4\Big] \text{sech}\left(\theta _1-\theta _2\right)
\end{align*}} & 0 \\[1ex]
$\overrightarrow{{\rm Tri}}^{gg^*}_3$ & \parbox[c]{11cm}{
\begin{align*}
& -2 \Big[8 \cosh 2\theta _1+3 \cosh 4\theta _1+8 \cosh 2\theta _2+ \\& +3 \cosh 4\theta _2-16 \cosh \left(\theta _1+\theta _2\right)+ \\& +8 \cosh \left(2 \left(\theta _1+\theta _2\right)\right)-4 \sinh ^2\left(\theta _1+\theta _2\right) \text{sech}^2\left(\theta _1-\theta _2\right)+10\Big]
\end{align*}} & 0 \\[1ex]
$\overrightarrow{{\rm Tri}}^{gg^*}_4$ & 0 & \parbox[c]{10cm}{
\begin{align*}
& -2 \Big[\cosh 4 \theta _1-4 \cosh 2\theta_2+\cosh 4 \theta _2 + \\& + 8 \cosh \left(\theta _1+\theta _2\right) -4 \cosh \left(3 \theta _1+\theta _2\right)+ \\& +4 \cosh 2\theta_1 \left(\cosh 2\theta_1 \text{sech}\left(\theta _1-\theta _2\right)-1\right)-10\Big]
\end{align*}} \\[1ex]
$\overrightarrow{{\rm Bub}}^{gg^*}_4$ & $32 \left(\cosh 2\theta_1+\cosh 2\theta_2-2 \cosh \left(\theta _1+\theta _2\right)\right)$ & 0 \\[1ex]
$\overrightarrow{{\rm Bub}}^{gg^*}_5$ & 0 & $-12 \left(\cosh 2\theta_1+\cosh 2\theta_2-2 \cosh \left(\theta _1+\theta _2\right)\right)$ 
\end{tabular}
\end{sideways}}
\caption{Table of coefficients for bubbles with momentum $p^2=-2$ for opposite helicity scattering and forward kinematics.}
\label{tab:forp}
\end{table}
\newpage
Finally, the tensor reduction of integrals with vanishing external momentum is spelled out in Table \ref{tab:opp0}.
\begin{table}[!ht]
\centering
\resizebox{!}{0.85\textheight}{
\begin{sideways}
\begin{tabular}{l|ccc}
 & ${\rm I}[2,2,0]$ & ${\rm I}[4,4,0]$ & ${\rm I}[1,1,0]$ \\[1ex]
 \hline\noalign{\smallskip\smallskip}
$\overrightarrow{{\rm Box}}^{gg^*}_b$ & $4 \left(\cosh \left(3 \theta _1+\theta _2\right)+\cosh \left(\theta _1+3 \theta _2\right)\right)$ & $-16 \left(\cosh 2\theta _1+\cosh 2\theta _2-2\right)$ & 0 \\[1ex]
$\overrightarrow{{\rm Box}}^{gg^*}_f$ & 0 & 0 & \parbox[c]{11cm}{
\begin{align*}
&
-4 \Big[1 + \cosh 2\theta _1 + \cosh 2\theta _2 + \\& 
+ \cosh \left(\theta _1+ 3 \theta _2\right)
+ \frac{\cosh ^2 2 \theta _1}{\cosh\left(\theta _1-\theta _2\right)}\Big]
\end{align*}} \\[1ex]
$\overrightarrow{{\rm Tri}}^{gg^*}_1$ & 0 & $16 \left(\cosh 2\theta _1+\cosh 2\theta _2-4\right)$ & 0 \\[1ex]
$\overrightarrow{{\rm Tri}}^{gg^*}_2$ & 0 & $-4 \left(\cosh 4\theta _1+\cosh 4\theta _2+2\right)$ & 0 \\[1ex]
$\overrightarrow{{\rm Tri}}^{gg^*}_3$ & 0 & 0 & 0 \\[1ex]
$\overrightarrow{{\rm Tri}}^{gg^*}_4$ & 0 & 0 & $2 \left(2 \cosh 2\theta _1+\cosh 4\theta _1+2 \cosh 2\theta _2+\cosh 4\theta _2+6\right)$ \\[1ex]
$\overrightarrow{{\rm Bub}}^{gg^*}_1$ & 0 & 32 & 0 \\[1ex]
$\overrightarrow{{\rm Bub}}^{gg^*}_2$ & 0 & 0 & 0 \\[1ex]
$\overrightarrow{{\rm Bub}}^{gg^*}_3$ & $2$ & $4$ & $-6$ \\[1ex]
\hline\noalign{\smallskip\smallskip}
total & $2 \left(2 \cosh \left(3 \theta _1+\theta _2\right)+2 \cosh \left(\theta _1+3 \theta _2\right)+1\right)$ & $-4 \left(\cosh 4\theta _1+\cosh 4\theta _2+1\right)$ & \parbox[c]{11cm}{
\begin{align*}
& \Big[ -2 \cosh 4\theta _1+\cosh \left(\theta _1-5 \theta _2\right)+2 \cosh \left(\theta _1-\theta _2\right) + \\&
+\cosh \left(5 \theta _1-\theta _2\right)-2 \cosh 4\theta _2-2 \cosh 2 \left(\theta _1+\theta _2\right) + \\&
+\cosh \left(3 \theta _1+\theta _2\right)+\cosh \left(\theta _1+3 \theta _2\right)-2\Big] \text{sech}\left(\theta _1-\theta _2\right)
\end{align*}}
\end{tabular}
\end{sideways}}
\caption{Table of coefficients for bubbles with momentum 0 for opposite helicity scattering and forward kinematics.}
\label{tab:opp0}
\end{table}

\subsection{Tensor reduction for opposite helicity backward scattering}

The tensor reduction of integrals with invariants $s$ and $u$ is summarized in Table \ref{tab:backpq}.
\begin{table}[!ht]
\centering
\resizebox{!}{0.85\textheight}{
\begin{sideways}
\begin{tabular}{l|ccc}
 & ${\rm I}[2,2;s]$ & ${\rm I}[4,4;s]$ & ${\rm I}[1,1;s]$ \\[1ex]
 \hline\noalign{\smallskip}
$\overleftarrow{{\rm Box}}^{gg^*}_b$ & $\frac{4 \cosh 2 \theta _1 \cosh 2 \theta _2}{\cosh\left(\theta _1-\theta _2\right)}$ & 8 $\left(\cosh \left(\theta _1+\theta _2\right)+2\right){}^2$ & 0 \\[1ex]
$\overleftarrow{{\rm Box}}^{gg^*}_f$ & 0 & 0 & $8 \sinh ^2\frac{\theta _1+\theta _2}{2} \left(\frac{\cosh \left(\theta _1+\theta _2\right)+1}{ \cosh\left(\theta _1-\theta _2\right)}+2\right)$ \\[1ex]
$\overleftarrow{{\rm Tri}}^{gg^*}_1$ & 0 & $-32 \left(\cosh \left(\theta _1+\theta _2\right)+2\right)$ & 0 \\[1ex]
$\overleftarrow{{\rm Tri}}^{gg^*}_2$ & 0 & $-8 \left(4 \cosh \left(\theta _1+\theta _2\right)+\cosh 2 \left(\theta _1+\theta _2\right)+1\right)$ & 0 \\[1ex]
$\overleftarrow{{\rm Tri}}^{gg^*}_3$ & $-\frac{8 \cosh 2 \theta _1 \cosh 2 \theta _2 \cosh ^2\frac{\theta _1-\theta _2}{2}}{ \cosh^2\left(\theta _1-\theta _2\right)}$ & 0 & 0 \\[1ex]
$\overleftarrow{{\rm Tri}}^{gg^*}_4$ & 0 & 0 & $-16 \sinh ^2\frac{\theta _1+\theta _2}{2} \left(\frac{\cosh \left(\theta _1+\theta _2\right)+1}{ \cosh\left(\theta _1-\theta _2\right)}+2\right)$ \\[1ex]
$\overleftarrow{{\rm Bub}}^{gg^*}_1$ & 0 & 32 & 0 \\[1ex]
$\overleftarrow{{\rm Bub}}^{gg^*}_2$ & 0 & $32 \cosh \left(\theta _1+\theta _2\right)$ & 0 \\[1ex]
$\overleftarrow{{\rm Bub}}^{gg^*}_3$ & $4\frac{\cosh 2 \theta _1 \cosh 2 \theta _2}{\cosh^2\left(\theta _1-\theta _2\right)}$ & $8 \cosh ^2\left(\theta _1+\theta _2\right)$ & $8 \sinh ^2\frac{\theta _1+\theta _2}{2} \left(\frac{\cosh \left(\theta _1+\theta _2\right)+1}{ \cosh\left(\theta _1-\theta _2\right)}+2\right)$ \\[1ex]
\hline\noalign{\smallskip}
total & 0 & 0 & 0\\[1ex]
\hline\hline\noalign{\smallskip}
 & ${\rm I}[2,2;u]$ & ${\rm I}[4,4;u]$ & ${\rm I}[1,1;u]$ \\[1ex]
 \hline\noalign{\smallskip}
$\overleftarrow{{\rm Box}}^{gg^*}_b$ & $-4 \frac{\cosh 2 \theta _1 \cosh 2 \theta _2}{ \cosh\left(\theta _1-\theta _2\right)}$ & $8 \left(\cosh \left(\theta _1+\theta _2\right)-2\right)^2$ & 0 \\[1ex]
$\overleftarrow{{\rm Box}}^{gg^*}_f$ & 0 & 0 & $-8 \cosh ^2\frac{\theta _1+\theta _2}{2} \left(2+\frac{\cosh \left(\theta _1+\theta _2\right)-1}{\cosh\left(\theta _1-\theta _2\right)}\right)$ \\[1ex]
$\overleftarrow{{\rm Tri}}^{gg^*}_1$ & 0 & $32 \left(\cosh \left(\theta _1+\theta _2\right)-2\right)$ & 0 \\[1ex]
$\overleftarrow{{\rm Tri}}^{gg^*}_2$ & 0 & $-8 \left(-4 \cosh \left(\theta _1+\theta _2\right)+\cosh 2 \left(\theta _1+\theta _2\right)+1\right)$ & 0 \\[1ex]
$\overleftarrow{{\rm Tri}}^{gg^*}_3$ & $\frac{8 \sinh ^2\frac{\theta _1-\theta _2}{2} \cosh 2 \theta _1 \cosh 2 \theta _2}{\cosh^2\left(\theta _1-\theta _2\right)}$ &  & 0 \\[1ex]
$\overleftarrow{{\rm Tri}}^{gg^*}_4$ & 0 & 0 & $16 \cosh ^2\frac{\theta _1+\theta _2}{2} \left(2+\frac{\cosh \left(\theta _1+\theta _2\right)-1}{\cosh\left(\theta _1-\theta _2\right)}\right)$ \\[1ex]
$\overleftarrow{{\rm Bub}}^{gg^*}_1$ & 0 & 32 & 0 \\[1ex]
$\overleftarrow{{\rm Bub}}^{gg^*}_2$ & 0 & $-32 \cosh \left(\theta _1+\theta _2\right)$ & 0 \\[1ex]
$\overleftarrow{{\rm Bub}}^{gg^*}_3$ & $4 \frac{\cosh 2 \theta _1 \cosh 2 \theta _2}{\cosh^2\left(\theta _1-\theta _2\right)}$ & $8 \cosh ^2\left(\theta _1+\theta _2\right)$ & $-8 \cosh ^2\frac{\theta _1+\theta _2}{2} \left(2+\frac{\cosh \left(\theta _1+\theta _2\right)-1}{\cosh\left(\theta _1-\theta _2\right)}\right)$ \\[1ex]
\hline\noalign{\smallskip}
total & 0 & 0 & 0 
\end{tabular}
\end{sideways}}
\caption{Table of coefficients for bubbles with momentum $s$ and $u$ for opposite helicity scattering and backward kinematics.}
\label{tab:backpq}
\end{table}
For integrals with external momentum inflowing, tensor reduction produces the results collected in Table \ref{tab:backp}.
\begin{table}[!ht]
\centering
\begin{tabular}{l|cc}
 & ${\rm I}[2,4;-2]$ & ${\rm I}[1,1;-2]$ \\[1ex]
 \hline\noalign{\smallskip}
$\overleftarrow{{\rm Box}}^{gg^*}_b$ & $-32 \sinh ^2\left(\theta _1+\theta _2\right)$ & 0 \\[1ex]
$\overleftarrow{{\rm Box}}^{gg^*}_f$ & 0 & $2\,\frac{\cosh 4 \theta _1 -4 \cosh 2\left( \theta _1 - \theta _2\right) + \cosh 4 \theta _2 - 2}{\cosh^2\left(\theta _1-\theta _2\right)}$ \\[1ex]
$\overleftarrow{{\rm Tri}}^{gg^*}_2$ & $32 \cosh 2 \left(\theta _1+\theta _2\right)$ & 0 \\[1ex]
$\overleftarrow{{\rm Tri}}^{gg^*}_3$ & $-8\,\frac{\cosh 4 \theta _1 +2 \cosh 2 \theta _1  \cosh 2 \theta _2 + \cosh 4 \theta _2 + 2}{\cosh^2\left(\theta _1-\theta _2\right)}$ & 0 \\[1ex]
$\overleftarrow{{\rm Tri}}^{gg^*}_4$ & 0 & 16 \\[1ex]
\hline\noalign{\smallskip}
total & $-4\,\frac{\cosh 4 \theta _1 + \cosh 4 \theta _2 + 2}{\cosh^2\left(\theta _1-\theta _2\right)}$ & $2\,\frac{\cosh 4\theta _1 + \cosh 4 \theta _2 +2}{\cosh^2\left(\theta _1-\theta _2\right)}$ 
\end{tabular}
\caption{Table of coefficients for bubbles with momentum $p^2=-2$ for opposite helicity scattering and backward kinematics.}
\label{tab:backp}
\end{table}

\section{Expanded Lagrangian}
\label{app:lagr_exp}

In this appendix we spell out the interaction terms of the Lagrangian \eqref{eq:lagrangian}, up to quartic order in the fields.
Cubic vertices read
\begin{align}
{\cal L}_3 &=
-4\tilde\phi\, |\partial_s x - x|^2 + 2 \phi [(\partial_t \phi)^2-(\partial_s \phi)^2] + 2 \phi\ [(\partial_t y^a)^2-(\partial_s y^a)^2] + \nonumber\\
&
+ 4i\, \phi [(\partial_s \bar\psi_i - \bar\psi_i) \Pi_{+} \psi^i + \bar\psi_i \Pi_{-} (\partial_s \psi^i - \psi^i)] + \nonumber\\
&
+ 2 i\, y^a [(\partial_s \bar\psi_i - \bar\psi_i) \P_{+} (\rho^{a6})^{i}_{\phantom{i}j} \psi^j - \bar\psi_i \Pi_{-} (\rho^{a6})^{i}_{\phantom{i}j} (\partial_s \psi^j - \psi^j) ] + 2i\, \partial_t y^a \bar\psi_i \gamma^t \P_{+} (\rho^{a6})^{i}_{\phantom{i}j} \psi^j +  \nonumber\\&
+ 2 (\partial_s x - x) (\psi^i)^T \P_{+} (\rho^6)_{ij} \psi^j - 2 (\partial_s x^* - x^*) \bar\psi_i \P_{-} (\rho^{\dagger}_6)^{ij} (\bar\psi_j)^T
\end{align}
and quartic interactions
\begin{align}
{\cal L}_4 &=
8\, \phi^2\, |\partial_s  x - x|^2 + 2\, \phi^2
[\partial_{\alpha} \phi \partial_{\alpha} \phi + \frac{2}{3} \phi^2]
+ 2 \phi^2 \partial_{\alpha} y^a\partial_{\alpha} y^a - \frac{1}{2}y^a y^a\, \partial_{\alpha} y^b \partial_{\alpha} y^b + \nonumber\\
&
- i (4\phi^2 -y^a y^a)\,[(\partial_s \bar\psi_i - \bar\psi_i) \Pi_{+} \psi^i + \bar\psi_i \Pi_{-} (\partial_s \psi^i - \psi^i)] + \nonumber\\&
- 4 i\, \phi\,y^a [(\partial_s \bar\psi_i - \bar\psi_i) \P_{+} (\rho^{a6})^{i}_{\phantom{i}j} \psi^j - \bar\psi_i \Pi_{-} (\rho^{a6})^{i}_{\phantom{i}j} (\partial_s \psi^j - \psi^j)] + \nonumber\\
&
- 6 \phi\,[(\partial_s x - x) (\psi^i)^T \P_{+} (\rho^6)_{ij} \psi^j - (\partial_s x^* - x^*) \bar\psi_i \P_{-} (\rho^{\dagger}_6)^{ij} (\bar\psi_j)^T] + \nonumber\\
&
+ 2(\partial_s x - x) (\psi^i)^T \P_{+} (\rho^a)_{ij} y^a \psi^j - 2(\partial_s x^* - x^*) \bar\psi_i \P_{-} (\rho^{\dagger}_a)^{ij} y^a (\bar\psi_j)^T + \nonumber\\
&
- 2 i\, y^a\partial_t y^b\, \bar\psi_i \gamma^t \P_{+} (\rho^{ab})^{i}_{\phantom{i}j} \psi^j + (\bar\psi_i \gamma^t \P_{+} (\rho^{a6})^{i}_{\phantom{i}j} \psi^j)^2 - (\bar\psi_i \gamma^t \P_{+} \psi^i)^2
\end{align}
In the computation of the first tadpole diagram a quintic vertex is needed
\begin{equation}\label{eq:phivertices}
{\cal L}^{x,\phi}_{5} = -\frac{32}{3}\, \phi^3\, \big| \partial_s x - x \big|^2
\end{equation}
\newpage

\bibliographystyle{JHEP-2}

\bibliography{biblio}

\end{document}